\documentclass[twocolumn]{aastex631}

\usepackage{longtable}

\shorttitle{Gaseous Counterrotating Galaxies in MaNGA}
\shortauthors{Beom et al.}

\graphicspath{{./}{figures/}}

\begin{document}

\title{SDSS IV MaNGA: The impact of the acquisition of gas with opposite angular momentum \\ 
on the evolution of galaxies}

\author[0000-0002-3032-2292]{Minje Beom}
\affiliation{Astronomy Department 
New Mexico State University 
Las Cruces, NM 88003, USA}

\author[0000-0002-0782-3064]{Ren\'e A. M. Walterbos}
\affiliation{Astronomy Department 
New Mexico State University 
Las Cruces, NM 88003, USA}

\author[0000-0002-3601-133X]{Dmitry Bizyaev}
\affiliation{Apache Point Observatory and New Mexico State University,
Sunspot, NM 88349, USA}
\affiliation{Sternberg Astronomical Institute, Moscow State University,
Universitetskiy prosp. 13, Moscow, 119234, Russia}

\begin{abstract}

A gaseous counter-rotating galaxy is a galaxy containing a gas component with opposite angular momentum to the main stellar disk. 
The counter-rotating gas provides direct evidence for the accretion of external material, a key aspect in hierarchical galaxy evolution. 
We identified 303 gaseous counterrotators out of 9992 galaxies in MaNGA. 
The majority of the counterrotators are early-types. 
This implies their formation is highly correlated with early-type galaxies although it is still difficult to know if one leads to the other. 
To disentangle which of the galaxy characteristics within a morphological class were changed by the accretion of counter-rotating gas, we carefully selected a comparison sample with similar fundamental galactic properties, but co-rotation in gas. 
This comparison shows that gaseous counter-rotation correlates with weak rotation in the stellar component, high central concentration of star forming regions, if present, and a higher fraction of central low ionization emission regions (cLIER) galaxies. 
The light distributions of the stellar components, dust and HI content (both low), and overall suppressed star formation rates are similar for both samples and seem typical for the morphological class. 
We claim that elliptical and about half of the lenticular counterrotators, those with weak rotation in the stellar component in the outskirts and central regions, likely have a major merger origin for the gas acquisition, and the other half of lenticulars, with stronger stellar rotation, may have a minor merger or pure gas accretion origin.

\end{abstract}


\section{Introduction} \label{sec:intro}

A counter-rotating galaxy is a galaxy containing a component 
with opposite angular momentum to the main stellar disk.
Early  studies have searched for counter-rotating components using 
long slit spectroscopic observations or 21-cm observations
\citep{Bertola1985,Galletta1987,BWK,Rix1992,Bertola1992}. 
The counter-rotating component can consist of stars or gas,
referred to as a stellar or gaseous counter-rotating galaxy, respectively
(gaseous counterrotators; \citet{Galletta1987,BWK,Bertola1992,Rix1995}, 
stellar counterrotators; \citet{Bertola1985,Rubin1992}, 
and see also many further references in the review papers of 
\citet{Galletta1996,BertolaCorsini1999,Corsini2014}).
Counterrotators have been reported to be relatively rare,
with higher fractions among early-type galaxies 
\citep{Bertola1992,Kuijken1996,KannappanFabricant2001,Pizzella2004}.
IFU (Integral Field Unit) surveys have allowed characterization of not only counterrotators, 
but also `misalignment galaxies' in which the stellar and gaseous disks
may be tilted by more than 30 degrees from each other.
$\rm ATLAS^{3D}$ found that a significant fraction of early-types were misalignment galaxies with low values of the spin parameter \citep{Davis2011}.
The SAMI (Sydney-AAO Multi-object Integral field Spectrograph) survey 
reported several tens of misalignment galaxies 
of all morphological types, with again a higher frequency in early-types \citep{Bryant2019,Ristea2022}.
In the Northern hemisphere, MaNGA (Mapping Nearby Galaxies at APO) 
led to the characterization of the morphological features and the global characteristics of 
hundreds of misalignment galaxies, see e.g. \citep{Jin2016,Li2021,Xu2022,Zhou2022}.
The observational studies of misalignment galaxies have focused on the comparison 
with aligned galaxies (i.e. corotators) to study their distinctive 
characteristics and how they formed in the context of the galaxy evolution
\citep{Duckworth2019,Xu2022,Zhou2022,Ristea2022}.
These recent IFU surveys, with their greatly enhanced sample sizes of misaligned galaxies,
enable detailed statistical analysis and characterization of their properties.
Simulation studies also help clarify how these systems may be formed
and how the misalignment angle may change over time \citep{Duckworth2020,Khim2020,Khim2021,Casanueva2022}.
These studies showed that the misalignment may be the result of
tidal interaction, environmental harassment, and massive outflows,
as well as mergers and gas accretion, and they
addressed their impact on the galaxy properties and evolution.
These observational and theoretical studies confirm that the acquisition of gas from an external source is 
an important process in galaxy evolution, particularly, it seems, among early-type galaxies.

A counter-rotating gas disk in a galaxy is the most direct evidence of gas acquisition from an external source. 
Not only that, it also provides information on how the gas collision between the pre-existing gas and the gas acquired can change galactic properties.
As \citet{Corsini2014},\citet{Chen2016}, and \citet{Jin2016} pointed out, 
gas acquisition with a large misalignment angle, i.e. for a counterrotator,
can result in large angular momentum loss through the gas collision.
In this paper, based on MaNGA, the largest IFU survey, 
we focus on gaseous counter-rotating galaxies
whose gaseous disk is misaligned by more than 150 degree to the stellar component.
We also carefully selected a comparison sample 
to have similar fundamental physical properties of the gaseous counterrotators
in terms of stellar mass, g-r color, but co-rotation of gas with stars.
In addition, we intended to distinguish the characteristics induced by the counter-rotating gas disk from the morphological characteristics.
Since the gaseous counterrotators are mostly early-types, 
their characteristics can be summarized by those of the early-type galaxies.
For example, our previous study, \citet{Beom2022} about edge-on disk counterrotators,
showed that their stellar component tend to be radially concentrated at the center, and
they are less abundant in gas and dust contents consequently with low star formation.
As we mentioned in the paper, it might be arisen from the morphological bias 
because the edge-on gaseous counterrotators resemble 
lenticulars, while many of the comparison sample looks like spirals.
To investigate the impact of the counter-rotating gas on the galactic properties
beside the morphological effects, 
we did a careful morphological classification, 
enabling comparisons within the same morphology.

In section 2, we describe the MaNGA IFU and archival data used in the analysis. 
In section 3, we introduce how the gaseous counterrotators and the comparison sample were selected and their morphological classification.
In section 4, we show the photometric and spectroscopic properties of the counter-rotating galaxies and of the comparison sample. 
In section 5, we discuss some unique cases among the gaseous counterrotators
that provide clues on their formation process.
In section 6, we will discuss how the physical properties of counter-rotating galaxies inform the possible formation and evolution of counterrotators.
In this paper, a Hubble constant of $70\ \rm km~s^{-1}~Mpc^{-1}$ $(h=0.7)$ is assumed.

\section{Data} \label{sec:Data}
\subsection{MaNGA} \label{sec:MaNGA_data}

We use the final version of the MaNGA (Mapping Nearby Galaxies at APO) data \citep{mangadr17}.
MaNGA is one of the survey projects in SDSS-IV (Sloan Digital Sky Survey)
using  Integral Field Unit (IFU) observations
to produce spatially resolved spectroscopic data for galaxies \citep{manga_overview,Blanton2017,Gunn2006,Drory2015,mangadr15,mangadr16,mangadr17}. 
MaNGA observed approximately 10,000 galaxies, 
selected as an unbiased sample in terms of 
stellar mass ($\rm M_{\star} > 10^{9}~M_{\odot}$)
\citep{Law2015,Wake2017,Yan2016a}. 
MaNGA provides spectra from 3600 to 10,300~\AA~ using 
the BOSS spectrographs with a spectral resolution of about 2100 at 6000~\AA~
\citep{BOSS,Yan2016b}. 
The observations were conducted to reach a S/N of $\rm 14$ to $\rm 35$ per spatial sample
in the stellar continuum at $\rm 1~R_{eff}$, 
which required about 3 hours net integration for each target.
The MaNGA targets were selected as two main subgroups and two minor subgroups 
\citep{manga_overview}.
The two main subgroups are called the primary sample and the secondary sample, 
separated by the IFU coverage of MaNGA.
The primary sample of about 5000 galaxies, a main subgroup, was selected 
to observe the central part out to $\rm 1.5~R_{eff}$. 
The secondary sample of about 3300 galaxies, another main subgroup, is  
observed by the IFU bundle out to $\rm 2.5~R_{eff}$. 
These samples of a wide range in stellar masses were selected without color bias.
The first minor subgroup is a `color-enhanced sample' of about 1,700 galaxies,
selected to include blue massive galaxies and green valley galaxies
for a study of the quenching process.
This `color-enhanced' sample is observed out to 
$\rm$ 1.5 $R_{eff}$, which is same as for the primary sample of MaNGA.
Another minor subgroup includes about 1000 ancillary targets, selected for various observations using the unique capability of the MaNGA instrument \citep{Yan2016a}.
The IFU bundles of the MaNGA consist of 19 to 127 fibers with hexagonal shape 
covering 12 to 32 $arcsec$ in diameter on sky. 
The spatial resolution of MaNGA is 2.5 $arcsec$, 
which corresponds to 1.3 to 4.5 $kpc$ for the primary sample and 
2.2 to 5.1 $kpc$ for the secondary sample. 
The redshift range is up to $0.15$, 
while about 60\% of the targets are within $0.02 \le z \le 0.05$, 
which corresponds to a few hundreds $Mpc$ distances.
We use all galaxy targets of MaNGA regardless of the main or minor groups. 
In the case that a galaxy was multiply observed, we took a deeper observation 
or an observation with wider IFU coverage.
As a result, 9992 galaxies\footnote{We did not count a group member of the target 
or a background galaxy observed within MaNGA IFU.} 
are selected as the total galaxy sample among 11,273 individual observations.

\subsection{Data Processing of MaNGA data} \label{sec:DAP_PIPE3D}
We use the data products from the MaNGA DRP 
(Data Reduction Pipeline, version 3.1.1) 
and DAP (Data Analysis Pipeline, version 3.1.0) \citep{DRP,DAP1,DAP2,DAP3}.
All figures and results are based on data with the default spaxel binning 
(spaxels are 0.5").
The velocity fields are taken from the results of DAP after the subtraction of the systemic velocity based on SDSS 3 arcsec spectroscopic redshift of each target. 
We adjusted a velocity field to have zero velocity at the center and
a symmetric rotational component on the disk/outskirts only in the case that 
a galaxy was found to have none-zero velocity at its center.
Most of the data were treated in the same way as \citet{Beom2022}, but with 
additional support from Pipe3D, a value added catalog (VAC) of MaNGA.
Pipe3D is a pipeline for IFU data that provides both the stellar population 
and the ionized gas properties \citep{pipe3D}.
The emission line strengths are estimated from integrated spectra 
of the central 2.5 $arcsec$ region using pPXF (Penalized Pixel-Fitting, \citet{pPXF}). 
The integrated galaxy spectra are summed from all valid spaxels 
where the mean surface brightness in g band or H$\alpha$ wavelength 
has S/N higher than $3$.

\subsection{Archival Data and MaNGA VAC} \label{sec:archieval_data}
We use additional archival data for photometric information and physical properties of galaxies.
The photometric properties are taken
from the NSA catalog (NASA-Sloan Atlas) and 
the WISE (Wide-field Infrared Survey Explorer) all-sky database \citep{NSA,WISE}.
The bulge-to-disk decomposition information is taken from \citet{Simard2011} and \citet{PyMorp}.
In addition, DECaLS images were used by us to check the galaxy morphology and search for nearby companions \citep{DECaLS}.
We also use MaNGA VACs of pipe3D, HI-MaNGA and GEMA (Galaxy Environment for MaNGA). 
HI-MaNGA survey is a project to ascertain
HI properties of the MaNGA targets using new observations by the
GBT (Green Bank Telescope) and archival observation data from
ALFALFA (The Arecibo Legacy Fast ALFA Survey) \citep{Haynes2018,Masters2019,Goddy2020,Stark2021}.

\section{Identification of the Gaseous Counterrotators and Comparison Sample Selection} \label{sec:Sample}

\subsection{Identification of the Gaseous Counterrotators} \label{sec:identification}

We define a gaseous counterrotators as a galaxy 
whose misalignment angle 
between stellar and gas components ($\Delta$ PA) is 
larger than 150 degree.
In our previous paper, we studied
edge-on counterrotators and showed that they have radially concentrated
distributions in star light and ionized gas emission, low dust contents,
and suppressed star formation rates \citep{Beom2022}. 
The edge-on perspective enabled us to directly observe the vertical extent and 
the misalignment angle relatively free from degeneracy between the intrinsic misalignment and 
the line of sight of the observation.
However, the edge-on perspective limits the number of galaxies
and analysis of trends with morphology.
In this study we identify gaseous counterrotators from 
the entire MaNGA galaxy sample, regardless of the inclination,
and we also classify the galaxy morphologies. The greatly enhanced sample size enables 
more reliable statistics of the frequency of counter-rotation with morphology type and analysis of common characteristics and variations between the objects.

\begin{figure*}[ht!]
\centering
\includegraphics[scale=0.90]{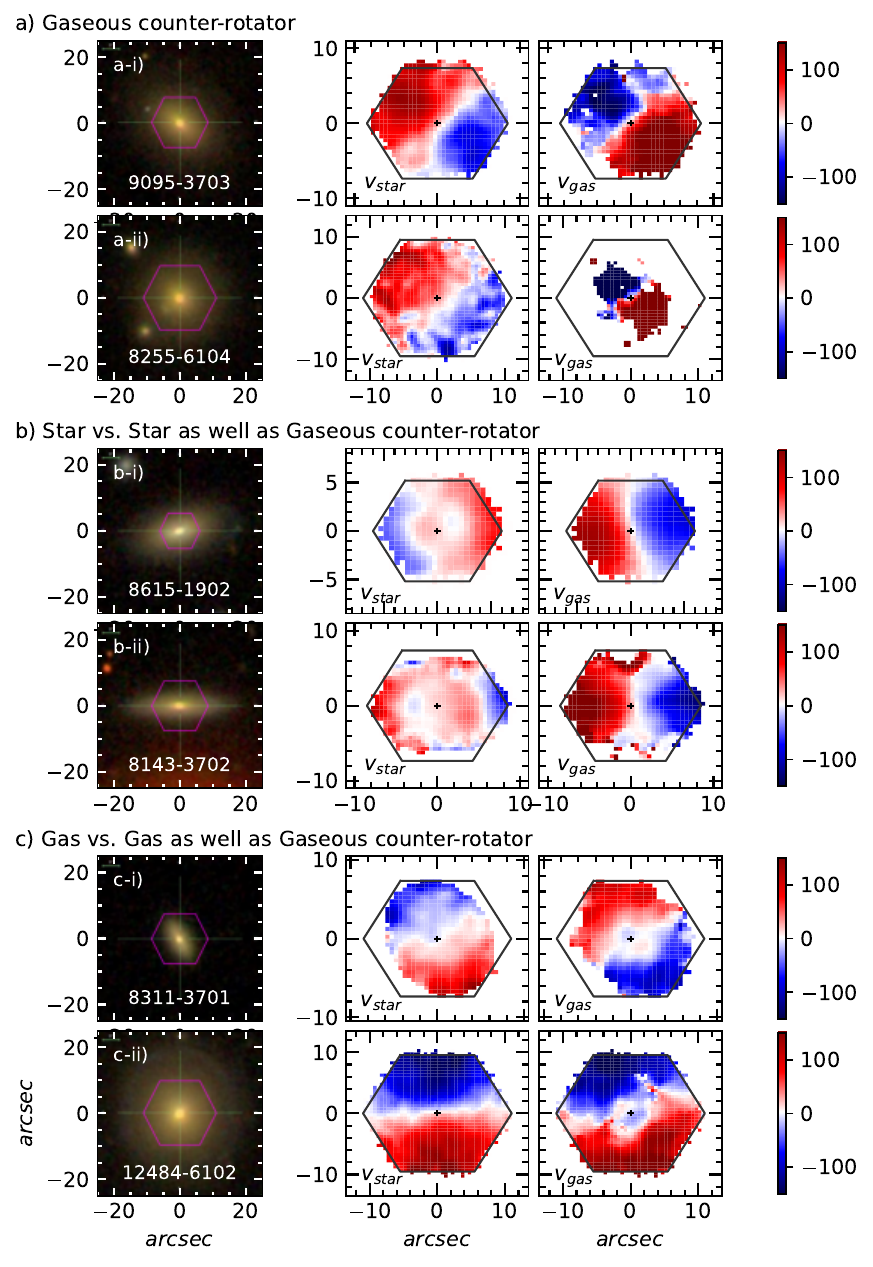}
\caption{Various examples of the gaseous counterrotators and their kinematics. 
 The purple hexagon in each left panel presents the coverage of the IFU 
for each target. It appears as a black hexagon on the velocity field map
in the middle and right panels.
The velocity map is in units of km s$^{-1}$.
The two cases in category a) are 
the majority of the gaseous counterrotators whose 
star and gas consist of  
one overall rotational component each. The middle two cases (b) are not only 
gaseous counterrotators but also stellar counterrotators
with the stellar kinematics showing two components 
rotating in opposite directions. 
The bottom two cases (c) are gas vs.~gas counterrotators
as well as gaseous counterrotators.
\label{fig:CRs_sample}}
\end{figure*}

Identification of the counterrotators was done by visual inspection
to detect any counter-rotating gas component,
regardless of their extent and location. 
Fig.~\ref{fig:CRs_sample} shows some examples of various gaseous 
counter-rotating galaxies. 
The upper two cases (a) are the majority of the sample. Their stellar and gaseous components show overall rotation in each component.
With overall rotation in this paper, we indicate
that a galactic component appears to have overall rotation over its extent.
The first case in Figure (a-i) is the most common of 
the counterrotators with the gas
component having a comparable size to the stellar component within the MaNGA IFU, 
while the second case (a-ii) has a gas component 
with overall rotation in a smaller central region. 
There are also infrequent (only a few) cases where
the overall rotation in the stellar component is smaller than 
that in the gas component within the MaNGA IFU.

The second set involves 29 stellar counterrotators identified from among 
the gaseous counterrotators.
These systems have two stellar components rotating in the opposite direction with one of them co-rotating with the gas component, i.e. \citet{Omori2021,Bevacqua2022}.
Case (b-i) in the figure shows the gas vs.~stars counter-rotation in the outer region,
but co-rotation at the center.
Case (b-ii) shows counter-rotation of the stars in the central region. 
A central stellar component counter-rotating to the other (main) stellar 
component may be too small to show its counter-rotation clearly. Such case may be confirmed by weak rotation (close to zero velocity)
with a central round shape.  Stellar counter-rotation in the central area with respect to the gas disk and outer stellar disk is strictly speaking not a case of gaseous counter rotation, however, such cases are rare in our sample (4 objects).
Case (b-i) is more frequent with 25 cases.
Note that 29 stellar counterrotators of cases (b-i) and (b-ii) 
include 9 and 8 objects of recent studies for stellar counterrotators in MaNGA \citet{Omori2021} and \citet{Bevacqua2022}, respectively.

The bottom two cases in Fig.~\ref{fig:CRs_sample} are 
gas vs.~gas counterrotators, 
i.e. cases where the gas components consist of two rotating components \citep{BWK,BureauChung,Chung2012}. 
As seen in panel b),
the counter-rotation between the gas vs.~ stars can be located 
at the outer region as in example (c-i), or at the center
as in example (c-ii).
The gas vs.~gas counterrotators are rare with only 7 cases confirmed; 
4 for (c-i) and 3 for (c-ii).

\begin{figure*}[ht!]
\centering
\includegraphics[scale=0.73]{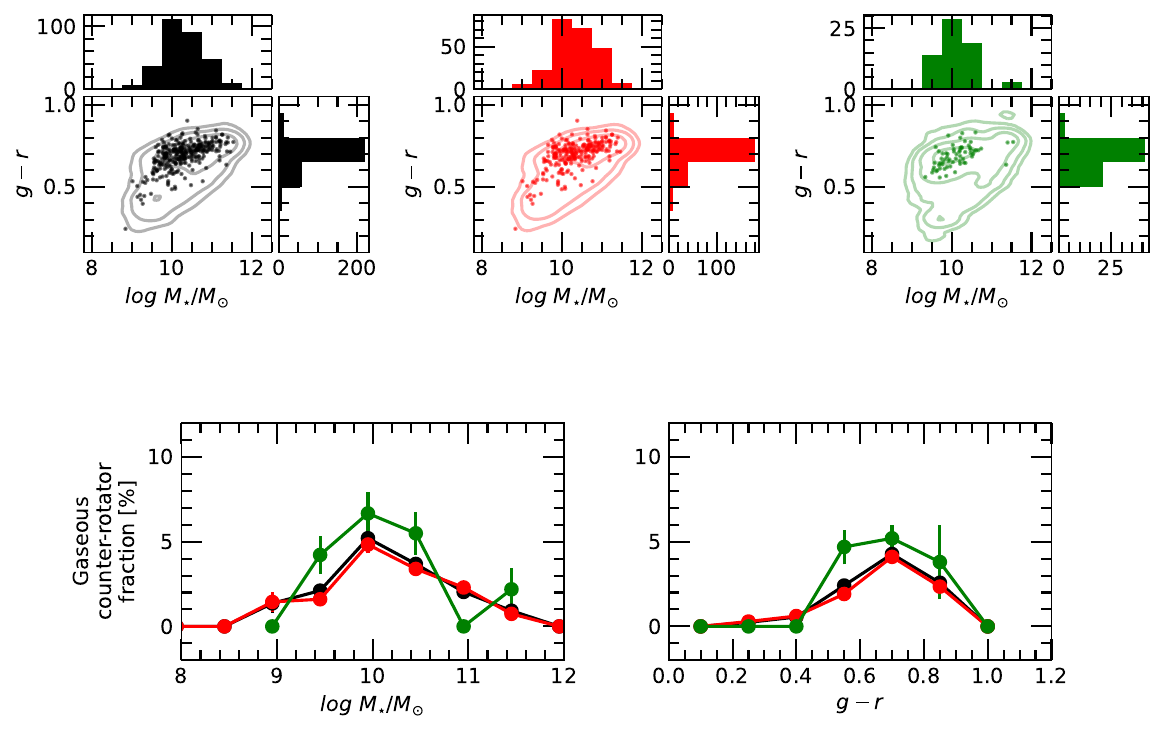}
\caption{The g-r color versus stellar mass diagram 
of the gaseous counterrotators, and their fractional frequency 
for each stellar mass and color bin within each subsample. 
The three colors indicate the subsample of MaNGA targets; 
all MaNGA galaxies (black), the `color-unbiased' sample (red), and 
the `color-biased' sample which includes more green valley galaxies (green). 
The dots and histograms show the distribution of the counterrotators 
on the color-mass diagram compared to the contours of the subsample.
The gaseous counterrotators are more frequent in the intermediate stellar mass range and the red sequence to the green valley regime on the diagram. \label{fig:hist}}
\end{figure*}

Adding all classes of counterrotators shown in Fig.~\ref{fig:CRs_sample}, we find 303 gaseous counterrotators, which corresponds
to 3.0\% of the whole galaxy sample of 9992. 
Fig.~\ref{fig:hist} shows the distribution of these objects 
on the g-r color and stellar mass diagrams for different samples,
and the fractions of the counterrotators
for each stellar mass and g-r color bin within the sample.
The primary sample and the secondary sample in the MaNGA sample were selected 
without color-bias, which is called a `color-unbiased' sample in this study (8466 galaxies). 
The central panels shows the distribution of the color-unbiased sample 
on the color-mass diagram as transparent red contours 
and the counterrotators in the sample as red dots and histograms.
The `color-enhanced sample' in MaNGA was selected to include more green valley galaxies and this sample is called `color-biased sample' in this paper (1526 galaxies).
The right panels shows the distribution of the color-biased sample as transparent green contours
and the counterrotators in the sample as the green dots and histograms.
The black symbols of the left panels show all the MaNGA galaxies (contours) and
all the counterrotators (dots and histograms).
The bottom two panels shows the fractions of the counterrotators
for each stellar mass and g-r color bin within the sample.
Regardless of color-biased or color-unbiased selection, 
the overall distribution is more concentrated near the
intermediate stellar mass range around log M$_{\star}$=10, and in
the green valley to the red sequence regime with g-r color from 0.5 to 0.8. 
The overall fraction of gaseous counterrotators is 2.8\% (238/8466) in the color-unbiased sample 
and 4.3\% (65/1526) in the `color-enhanced sample' which includes more green valley galaxies.
This also confirms that the counterrotators are more likely to be 
a `green valley galaxy'.

\subsection{The Comparison Sample} \label{sec:comparison}

We selected a comparison sample to have fundamental galaxy properties 
similar to those of the counterrotators, but co-rotation between the stellar and gas components. 
In the first step, we narrow down candidates for each counterrotator using criteria of 1) inclination ($\Delta$ b/a $<$ $\pm$ 0.1), 2) IFU coverage (same IFU bundle in the observation and relative difference of the distance less than 20\%), 3) presence of overall rotation in both stellar and gaseous components and 4) their misalignment angle less than 30 degrees for the co-rotation.
In the next step, we selected up to five galaxies in the order of the closeness to the counterrotator on the color-stellar mass diagram. Also, we selected a comparison galaxy only the inside of the box of $\Delta$ (g-r) $<$ $\pm$ 0.1, and 0.5 $<$ M$_{\star,comparison}$/M$_{\star,CR}$ $<$ 2.0 on the diagram. 
Through these two steps, we selected at least two and up to five comparison galaxies with similar g-r color, stellar mass, inclination and IFU coverage 
to those for each counterrotator.
After eliminating overlap, the comparison sample consisted of about 600 galaxies.
In the final step, we excluded from the comparison sample those galaxies showing galaxy interaction features or a fore/background star/galaxy within the IFU field.
This leaves a final comparison sample of 558 galaxies.
Our selection procedure ensures that the comparison sample, by design, 
is biased in the same way as the gaseous counterrotator:
similar color range, intermediate stellar mass range, sufficient ionized gas to have a velocity field mapped by MaNGA,  
and rotation support in its kinematics.

\subsection{Morphological Classification} \label{sec:Morp}

We took the morphological classification from \citet{Vazquez-Mata2022}.
We labeled ellipticals (E) in this paper as those with -5 in their T-type classification. 
The galaxies with 0 in their T-type  are classified as lenticulars (S0).
We considered galaxies with -2  in their T-type as galaxies somewhat uncertain between the ellipticals and the lenticulars, 
and refer to them as `E/S0’ in this paper.
We also classified spirals as galaxies with 1 to 7 in their T-type classification. 
Since there is morphological ambiguity for highly inclined galaxies, 
we chose to classify edge-on disk (i.e. lenticular and spiral) galaxies as S$_{unc}$ rather than
adopt the classification provided by \citet{Vazquez-Mata2022} for such edge-on galaxies. 
As a final step, we added another separate category in the classification for recent or ongoing interacting galaxies. These are 
identified by a disturbed feature, a contact feature with a companion galaxy,
or a kinematic alignment of the companion to the counter-rotating gas.
More details about this classification are in Section \ref{sec:Mergers}.
We grouped these merger candidates as `Mergers' (M).
We do not list morphological classification for them since they tend to have asymmetric or disturbed morphology.

Any morphological classification is subjective and somewhat uncertain. 
We confirmed the main results and conclusions 
in this study using our own morphological classification. 
We found that the main results and conclusions based on 
the statistical analysis (average properties within each morphological classification) remain valid,
even though we had some disagreement in the morphological classification between ours and theirs. Here we present only the results using the morphological classification of \citet{Vazquez-Mata2022} 
to remove any potential bias in classification from our side.
Since early-type galaxies\footnote{In this paper, `early-types' include E, E/S0, and S0.} are dominant among the gaseous counterrotators (about 80\%),
we focus on their physical properties compared to the early-types in the comparison sample (corotators).
The comparison sample has spirals at a similar fraction (50.05\%) to  early-type galaxies (49.95\%), even though 
we did not use morphological criteria in their sample selection and 
most of them were selected to have colors corresponding to the red sequence and green valley regimes on the CMD.

\begin{deluxetable*}{ccccccccc}
\tablenum{1}
\tablecaption{
The frequency of counterrotators and comparison galaxies by morphology and emission line classification and star formation rates and HI detection rates
\label{tab:morp}}
\tablewidth{0pt}
\tablehead{
\colhead{Sample} & \colhead{Morphology} & \colhead{\# of Galaxies}& 
\multicolumn4c{Emission Line Diagnostic Class}& \colhead{$log$ $SFR/SFR_{SFMS}$} & \colhead{HI-MaNGA}\\ 
\nocolhead{} & \nocolhead{} & \nocolhead{}& 
\colhead{AGN}& \colhead{cLIER}& \colhead{cSF}& \colhead{N/A}& \nocolhead{} & \colhead{Detection Rate}
}
\startdata
    & E     & 49 &      4 (8) & 20 (41) & 9 (18) & 16 (33) &        
    -1.89$^{+0.74}_{-0.56}$ & 8/35 (23)\\
    & E/S0  & 68 &      6 (9)& 24 (35)& 20 (29) & 18 (26) &      
    -1.86$^{+0.92}_{-0.53}$ & 16/53 (30)\\
CRs & S0    & 40 &     7 (18)& 10 (25)& 16 (40)& 7 (18)&     
    -1.64$^{+0.72}_{-0.73}$ & 7/34 (21)\\
    & Sp    & 48 &       7 (15) & 11 (23)& 24 (50)& 6 (12)&         
    -1.49$^{+0.78}_{-0.78}$ & 22/47 (47)\\
    & S$_{unc}$& 65 &    10 (15)& 23 (35)& 24 (37)& 8 (12)&      
    -1.69$^{+0.82}_{-0.74}$ & 11/60 (18)\\
\hline
     & E     & 52 &    8 (15)& 18 (35)& 10 (19)& 16 (31)&        
     -1.97$^{+0.53}_{-0.61}$ & 6/32 (18)\\
     & E/S0  & 89 &    12 (13)& 21 (24)& 34 (38)& 22 (25)&      
     -1.73$^{+0.79}_{-0.60}$ & 18/68 (26)\\
com. & S0    & 69 &    7 (10)& 19 (28)& 27 (39)& 16 (23)&     
     -1.62$^{+0.68}_{-0.70}$ & 13/50 (26)\\
     & Sp    & 219 &   9 (4)& 37 (17)& 148 (68)& 25 (11)&      
     -1.08$^{+0.58}_{-0.70}$ & 85/171 (49)\\
     & S$_{unc}$& 129 &  8 (6)& 29 (22)& 77 (60)& 15 (12)&      
     -1.45$^{+0.65}_{-0.79}$ & 40/112 (35)\\
\hline
CRs  & Early-types & 157 &      17 (11)& 54 (34)& 45 (29)& 41 (26)&   
-1.93$^{+1.11}_{-0.47}$ & 31/122 (25)\\
com. & Early-types & 210 &      27 (13) & 58 (28)& 71 (34)& 54 (26)&  
-1.71$^{+0.82}_{-0.65}$ & 37/150 (24)\\
\hline
CRs  & Mergers & 33 &           5 (15)& 14 (42)& 7 (21)& 7 (21)& 
-1.84$^{+0.87}_{-0.53}$ & 7/19 (37)\\
\enddata
\tablecomments{The emission line ratio diagnostic classification is based on the ratios of [SII]/H$\alpha$ and [OIII]/H$\beta$, and the classification lines taken from \citet{BPTlines}.
The emission lines were measured in the central 2.5 arcsec and the classification 
is shown as 
active galactic nucleus (AGN),
central low ionization emission regions (cLIER), 
central star forming region (cSF), and 
galaxies whose one or more major emission line were not detected (N/A).
The numbers in the parentheses are the percentages in each morphological group.
The star formation rate (SFR) is present as the value compared to that of the star forming main sequence (SFMS) with the uncertainty estimated by 16th and 84th percentile values. 
This is defined as the deviation from the SFR predicted from the SFMS, on a log scale ($\Delta$SFR).
The detection rate of HI-MaNGA is present as the number of galaxies detected/observed with the detection percentage in parenthesis.}
\end{deluxetable*}

The morphological classification for the gaseous counterrotators 
results in 49 E, 68 E/S0, 40 S0, 48 Sp, 65 S$_{unc}$ and 33 Mergers.
The morphological classification is summarized 
in Table~\ref{tab:morp}
with other information introduced in later sections.
The morphological classification of the comparison sample results in 
52, 89, 69, 219, and 129 from E, E/S0, S0, Sp, to S$_{unc}$.
The comparison sample does not have the merger category because 
galaxies showing merger evidence were excluded in the sample section.
In contrast to the counterrotators, 
about a half of the comparison sample are spiral galaxies, 
including 38 late-type spirals.

We note that `ellipticals' (E and E/S0), as defined in this paper, are somewhat different from
traditional `ellipticals' in galaxy classification. 
Since we selected the gaseous counterrotators and the comparison sample as galaxies
where both the stellar and gaseous components have an overall rotation in each,
both the counterrotators and comparison sample are biased towards gas `rich' and `rotationally supported' systems.
Thus, `ellipticals' in this paper would mean relatively gas rich and rotational supported ellipticals.
Also, spiral galaxies in the comparison sample are different from traditional `spirals' because they are red spirals that were selected with similar g-r as the gaseous counterrotators.

Inspired by the fact that early-types are dominant in the gaseous counterrotators, 
we estimate the fraction of the counterrotators only in early-types.
The early-type counterrotators consist of about 5.58\% (157/3712) 
among 3712 early-types in MaNGA galaxies.
If we estimate the fraction using only the early-type with the overall rotation in \emph{both}
stellar and gaseous components, 
the counterrotators consist of 12.6\% (207/1644).
We will discuss these fractions and the possible implications regarding their formation in section \ref{sec:Discussion}.

\section{Physical Properties} \label{sec:properties}

In the following sections we discuss various physical properties of the counterrotators and the comparison sample of corotators, separated by morphological class. 
We note that the sample sizes of the counterrotators and the comparison sample 
for ellipticals (E and E/S0) and lenticulars (S0)
are large enough for comparison. 
We will mainly focus on the properties 
of the early-types in each sample
because they are dominant among the counterrotators.
Given the morphological uncertainty,
S$_{unc}$ is also excluded in the comparison of Section \ref{sec:BPT} to \ref{sec:groups}.
As described in the previous section, care should be taken when comparing the two samples for the class marked S$_{unc}$ since the comparison sample likely includes a higher fraction of spirals in that class.

\subsection{Optical Photometric Properties} \label{sec:Optical}

\begin{figure*}
\centering
\includegraphics[scale=0.5]{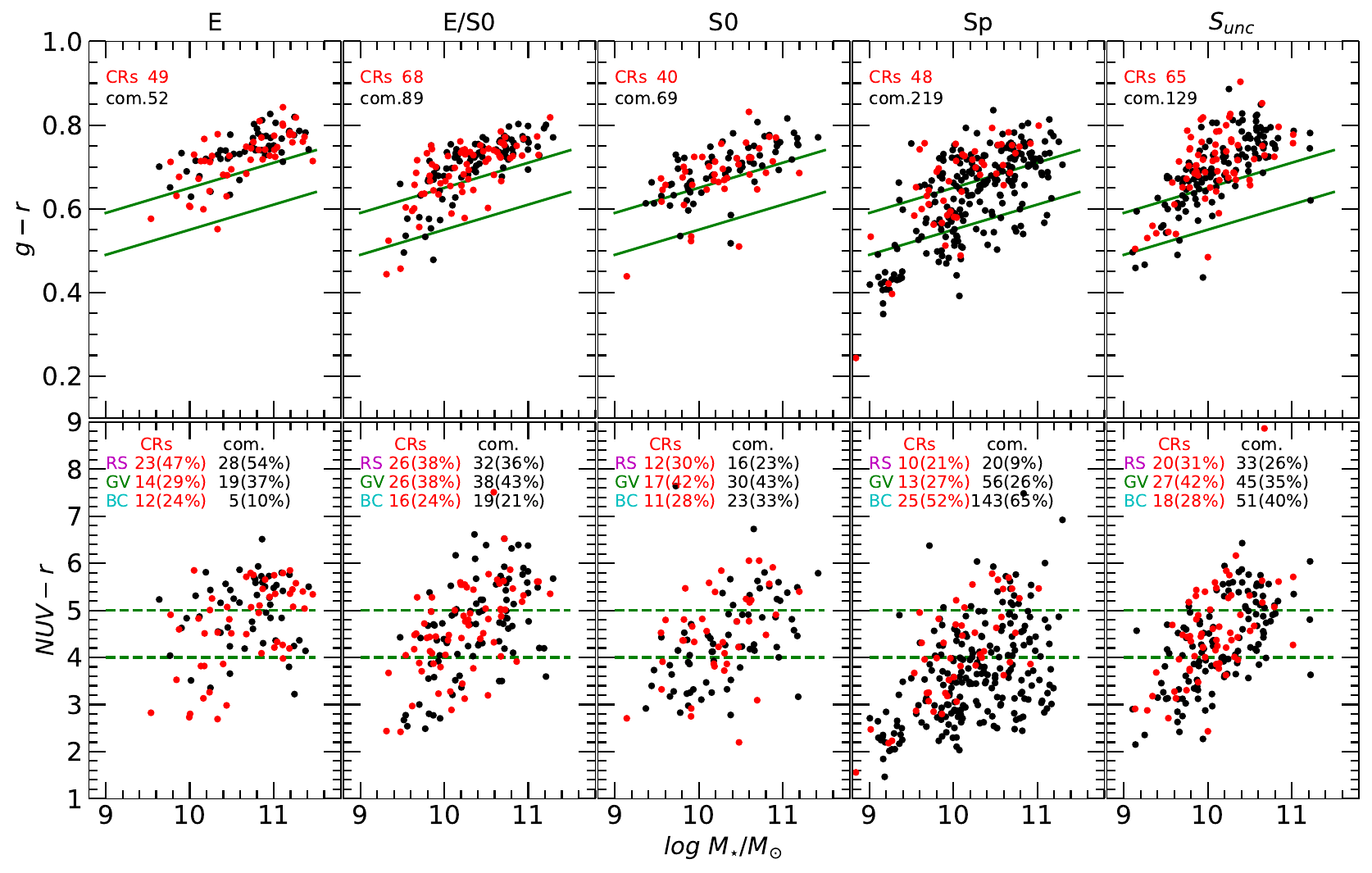}
\caption{The color versus stellar mass diagrams separated into morphological classes. 
The upper panels show g-r color and the bottom panels do NUV-r.
Each panel shows the counterrotators with red dots and the comparison sample with black dots.  The morphological types are: ellipticals (E), somewhat uncertain between E and S0 (E/S0), lenticulars (S0), spirals (Sp), and high inclination disk galaxies (S$_{unc}$). 
The green solid and dashed lines present 
the green valley (GV) regime on each g-r and NUV-r color, which are taken from \citet{LacknerGunn2012,Belfiore2018}.
The red sequence (RS) and the blue cloud (BC) galaxies are counted 
depending on the location on diagram; redder or bluer than the GV regime.
The numbers in the upper panels
present the number of galaxies in each morphology,
and the numbers in the bottom panels
are the number of galaxies depending on the color classes 
with their fractions within each morphology in the parenthesis.
\label{fig:CMD_morp}}
\end{figure*}

Fig.~\ref{fig:CMD_morp} shows the broadband color versus stellar mass diagrams
of the gaseous counterrotators and the comparison sample. 
The comparison sample was selected to have similar g-r color and stellar mass
so that the overall distributions of both samples
appear similar to each other on the diagrams.
Both samples are mostly in the red sequence, and 
some of them are on the green valley regime of 
the traditional g-r versus stellar mass diagram, 
the upper panels of Fig.~\ref{fig:CMD_morp}.
The NUV-r color diagram, the bottom panels of Fig.~\ref{fig:CMD_morp},
is more sensitive to the current and recent star formation 
so that it is better to see if the recent star formation still occurs.
In the NUV-r color classification for the early-types,
both samples includes somewhat many (up to about 30\%) galaxies showing the recent star formation. 
This is the sample selection effect that galaxies of both samples have enough gas to show overall rotation.
The green valley and the red sequence galaxies are about 35\% 
in both counterrotators and comparison samples.
A noticeable characteristic is that 
elliptical counterrotators are more on the blue cloud regime of Fig.~\ref{fig:CMD_morp}
than elliptical comparison galaxies,
while lenticular counterrotators are more on the red sequence regime on CMD
than  lenticular comparison galaxies.
This might imply that 
elliptical counterrotators preferentially have more recent star formation 
while lenticular counterrotators preferentially have less recent star formation
compared to each morphological class of the comparison sample.
Within the spirals of both samples, 
the comparison samples are more concentrated toward the blue cloud regime of Fig.~\ref{fig:CMD_morp}.
Since S$_{unc}$ in the comparison sample likely includes more spirals,
they, too, preferentially have more recent star formation 
than the counterrotators in that class.

\begin{figure*}[ht!]
\centering
\includegraphics[scale=0.55]{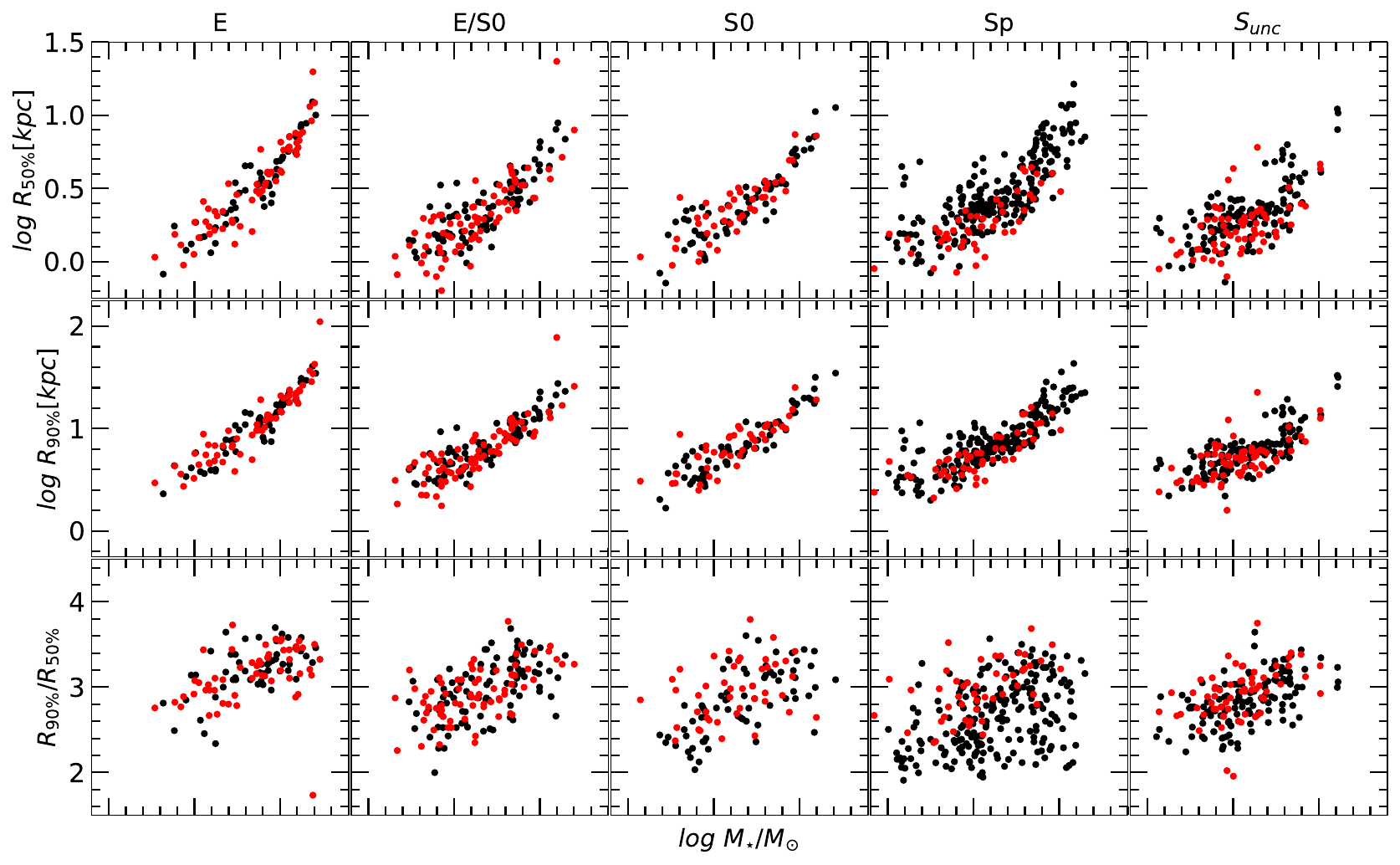}
\caption{The effective and 90\% light radii and their ratio, a proxy of the concentration index.
The value are taken from the petrosian radii in \citet{NSA}.
The symbols and the morphological classes for each panel are the same as in Fig.~\ref{fig:CMD_morp}.
There is no noticeable difference in the size and the concentration 
of the stellar component between the counterrotators and the comparison sample.
\label{fig:ER}}
\end{figure*}

In Fig.~\ref{fig:ER}, we plot the effective and 90\% light radii, and their ratios (a measure of the concentration index)
versus the stellar masses for the counterrotators and the comparison sample.
Within each morphology of E, E/S0, or S0, 
we find no difference in 
the size-mass distributions and the concentration index 
between the counterrotators and the comparison sample.
Within each morphology of Sp, or S$_{unc}$, on the other hand, 
the counterrotators are preferentially smaller 
and more concentrated at the center than the comparison samples. 
Their small effective radius leads to high concentration index values.

The result that the early-type counterrotators
have no significant size difference from the early-type comparison samples
may seem to contradict our previous result in \citet{Beom2022} 
that the counterrotators are smaller and have high concentration 
in the stellar component compared to other galaxies.
The previous study analyzed edge-on disk counterrotators
compared to other edge-on disk galaxies, including spirals as well as lenticulars. 
The rightmost column panel of Fig,~\ref{fig:ER} in this paper
is close to the representation of the previous result that the counterrotators appear to be small and more concentrated compared to the comparison sample.
We emphasize again, however, this is because the
S$_{unc}$ comparison sample in this study (the edge-on disk galaxies, non-counterrotator in the previous study) includes many spirals. 
So, our result in this study shows that
the result about the size and the concentration in the stellar component 
arose from a morphological effect.
Within each morphology, we also compared the S\'ersic index and the B/T ratios 
between the counterrotators and the comparison sample,
which show no noticeable difference as well.

\subsection{Dust and HI Content} \label{sec:dust}

\begin{figure*}[ht!]
\centering
\includegraphics[scale=0.50]{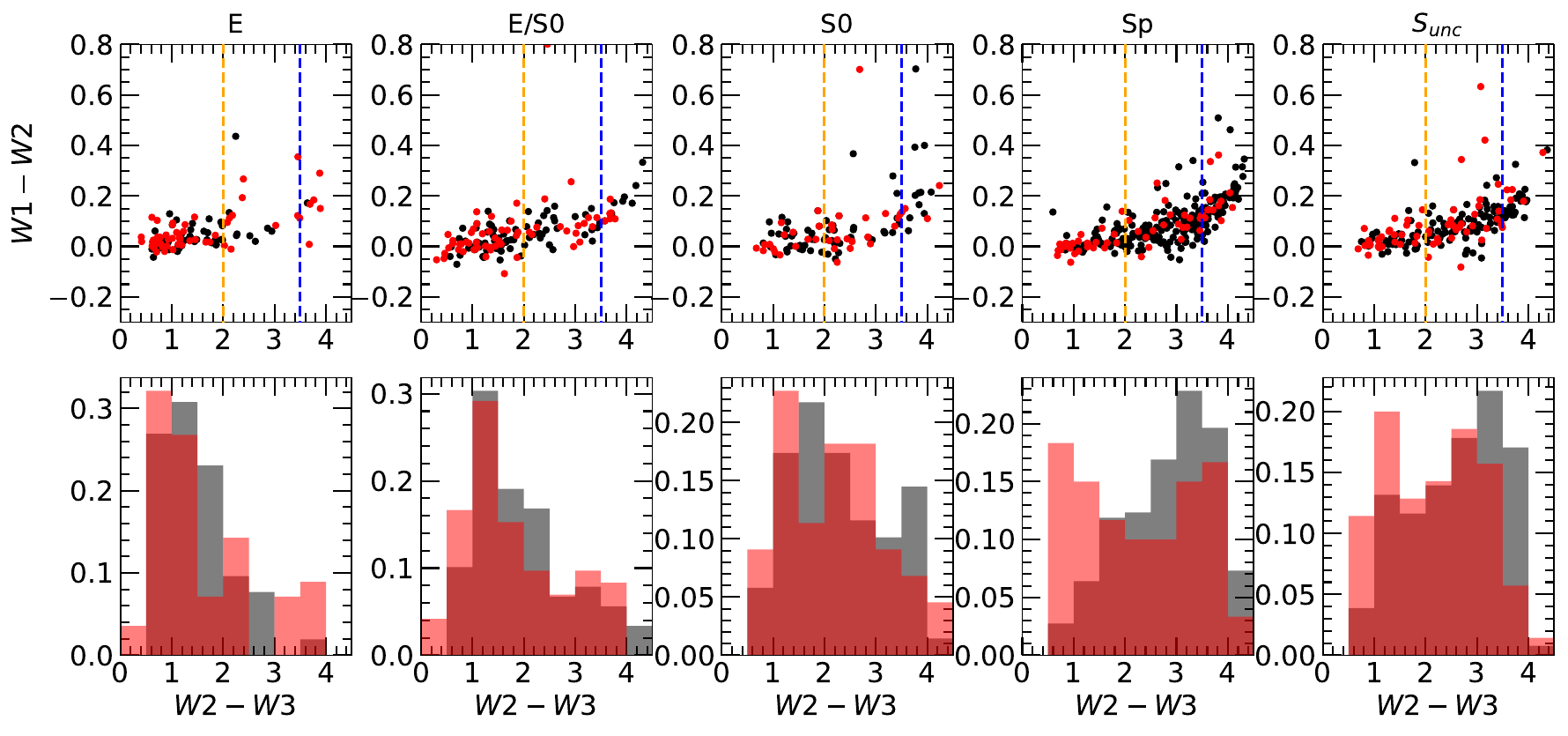}
\caption{The color-color diagram using W1, W2, W3 magnitudes from
WISE infrared observations, and the histogram of W2-W3 color. 
A high value of W2-W3 indicates high dust content.
The lines indicating classes are taken from \citet{Jarrett2013}; 
left from the yellow dashed line for mostly Spheroids, 
right from the blue dashed line mostly for star forming disks, and
the regime between them for intermediate types.
The symbols and the morphological order of each panel 
are the same as in Fig.~\ref{fig:CMD_morp}. 
The counterrotators are shown in red and the comparison sample is shown in grey in the histograms. 
The counterrotators looks to have somewhat lower dust content, which is insignificant based on KS test. 
\label{fig:wise}}
\end{figure*}

We analyzed the dust content using the WISE color W2-W3
for the counterrotator and the control samples.
Fig.~\ref{fig:wise} shows the WISE color-color diagrams for W2-W3 and W1-W3
and the histogram of the W2-W3 color distribution.
The higher the W2-W3 color a galaxy is, the more abundant dust content the galaxy has \citep{Jarrett2011,Jarrett2013}.
The figure includes only galaxies whose WISE magnitudes are detected; 
91\% of the counterrotators and 97\% of the comparison sample are plotted.
As seen in the figure, the counterrotators in E, E/S0, S0 
appear to have dust abundance similar to that of the comparison sample.
We found slightly more dust-poor galaxies whose W2-W3 color is lower than 2 
in the early-type counterrotators compared to those of the early-type comparison samples. 
However, the difference is not significant in a KS test.
A noticeable characteristic is that 
the distribution of W2-W3 values of the spirals of the counterrotators 
appears as the sum of dust-poor and dust-rich populations.
On the other hand, the spirals in the comparison sample are distributed toward high dust abundance as one population whose W2-W3 color peaks around 3.
Since S$_{unc}$ consists of lenticulars and spirals, 
W2-W3 values of both the counterrotators and the comparison samples
look similar to the sum of dust-poor (S0) and dust-rich (Sp) galaxies.

The abundance of neutral hydrogen gas (HI) is also compared 
between the counterrotators and the comparison sample.
The HI observations are taken from HI-MaNGA of \citet{Stark2021}. 
Since only galaxies with relatively abundant HI gas were detected,
the detection rates, summarized in the table \ref{tab:morp}, 
are more meaningful than the HI mass detected.
The overall detection rate is 27\% in the counterrotators, lower than the 37\% in the comparison sample. This may be accounted for by the morphological bias, in that the comparison sample includes many spiral galaxies.
If we restrict both samples to early-type galaxies, the detection rates are similar:
25.4\% in the counterrotators and 24.7\% in the comparison sample.
Within spirals, the detection rates are 43\% in the counterrotators and
44\% in the comparison samples.

\subsection{The Spin Parameter} \label{sec:spin}

The spin parameter is taken to be a proxy of how strong the galaxy rotation is to the velocity dispersion. 
It is estimated by the following equation \citep{Emsellem2007}.
\begin{equation}
\lambda_{1R_{e}} =  \frac{\sum_{i}{F_{i}R_{i}v_{i}}} {\sum_{i}F_{i}R_{i}\sqrt{{v_{i}^{2}+\sigma_{i}^{2}}}}
\end{equation}
where $F_{i}$, $v_{i}$ and $\sigma_{i}$ are the surface brightness, 
the line-of-sight velocity, and the velocity dispersion 
of the stellar component within the $i$ spaxel of the IFU data. 
The parameter $\lambda_{1R_{e}}$ is estimated from the spaxels enclosed by 
the isophotal ellipse at the effective radius.

\begin{figure*}[ht!]
\centering
\includegraphics[scale=0.50]{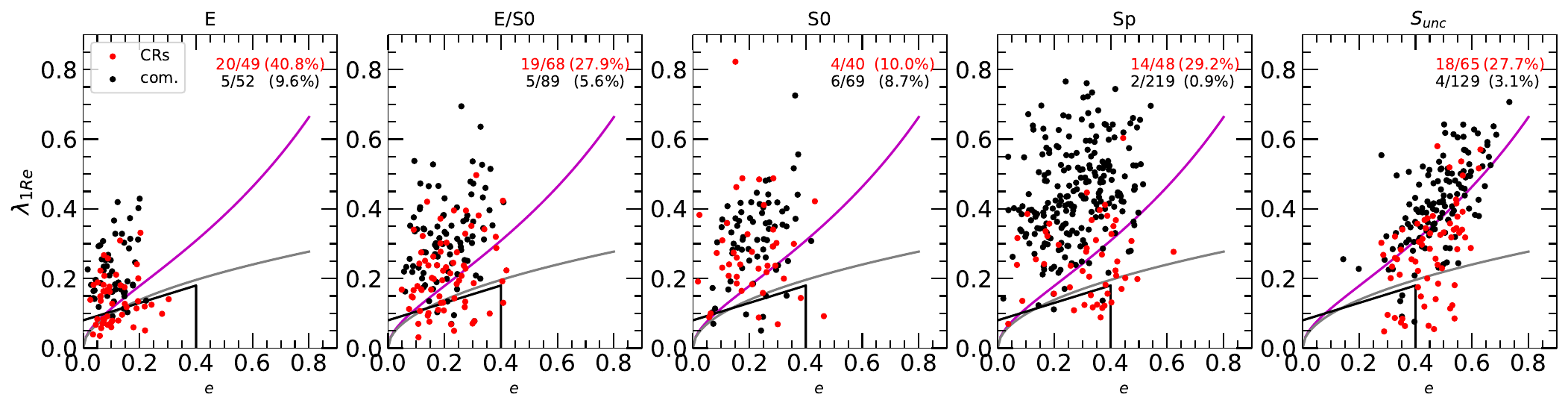}
\caption{The spin parameters of the counterrotators and the comparison sample for each morphological class.
The spin parameter is estimated for the stellar component within 1 $R_{eff}$.
The symbols and the morphological order of each panel 
are the same as in Fig.~\ref{fig:CMD_morp}.
The slow rotators are galaxies located in the bottom regime under the grey line taken from \citep{Emsellem2011}. 
The fractions of the slow rotators are presented in the top right corner
of each panel.
The counterrotators have low spin parameter compared to the comparison sample.
\label{fig:spin}}
\end{figure*}

Fig.~\ref{fig:spin} shows the spin parameters of the counterrotators and 
the comparison sample for each morphology. 
For all morphologies, the counterrotators have 
lower values of the spin parameter compared to the comparison sample.
This agrees with the result that misalignment galaxies have low spin parameter values 
\citet{Xu2022,Zhou2022,Ristea2022}.
Also, the high fraction of slow rotators among the gaseous counterrotators is in the agreement with that stellar counterrotators are frequently identified as slow rotators on the result of \citet{Emsellem2011}. 
It is surprising that even though the spin parameter is derived for the stars, 
the gaseous counterrotators have lower spin parameters. 
This may suggest the formation process of gaseous counterrotators 
can result in a significant effect on the kinematics in the stellar component.
We will discuss the details in the section \ref{sec:discussion_spin_parameter}.

A slow rotator is defined as a early-type galaxy to have weak overall rotation 
compared to its velocity dispersion, which appears to be lower than the gray line on Fig.~\ref{fig:spin} \citep{Emsellem2011}. 
The fraction of the slow rotators for each morphology 
is presented in the top right corner of each panel. 
We confirm that
counterrotators are more frequently identified as slow rotators
as well as the morphological trend that 
early-types are less rotationally supported systems than spirals.
The slow rotators consist of 27.4\% in the early-type counterrotators (28.1\% in the whole sample of counterrotators).
This is significantly higher than 7.62\%
in the early-type comparison sample (3.9\% in the whole comparison sample).
To be specific, the fraction of the slow rotators is calculated to be
40.8, 27.9, and 10.0\% in E, E/S0, and S0 for the counterrotators, respectively, versus 9.6, 5.6, and 8.7\% for the comparison sample.
Within spirals, the fractions of the slow rotators
are 29.2\% in the counterrotators, and 0.9\% in the comparison samples.
The fraction of slow rotators is significantly higher in the counterrotators than that of the comparison sample for all the morphological bins. 
One possible cause is that 
counterrotators host co- and counter-rotating stars. This can potentially be identified and disentangled by more detailed spatial and spectral decomposition of the light and kinematics data (i.e. \citet{Rubino2021} and \citet{Bevacqua2022}).
Another possible mechanism is that the weak overall rotation stems from their formation process. 
In Section \ref{sec:discussion_spin_parameter}, we will discuss these in detail.

\subsection{The Emission Line Ratios in the Central Region} \label{sec:BPT}

\begin{figure*}[ht!]
\centering
\includegraphics[scale=0.50]{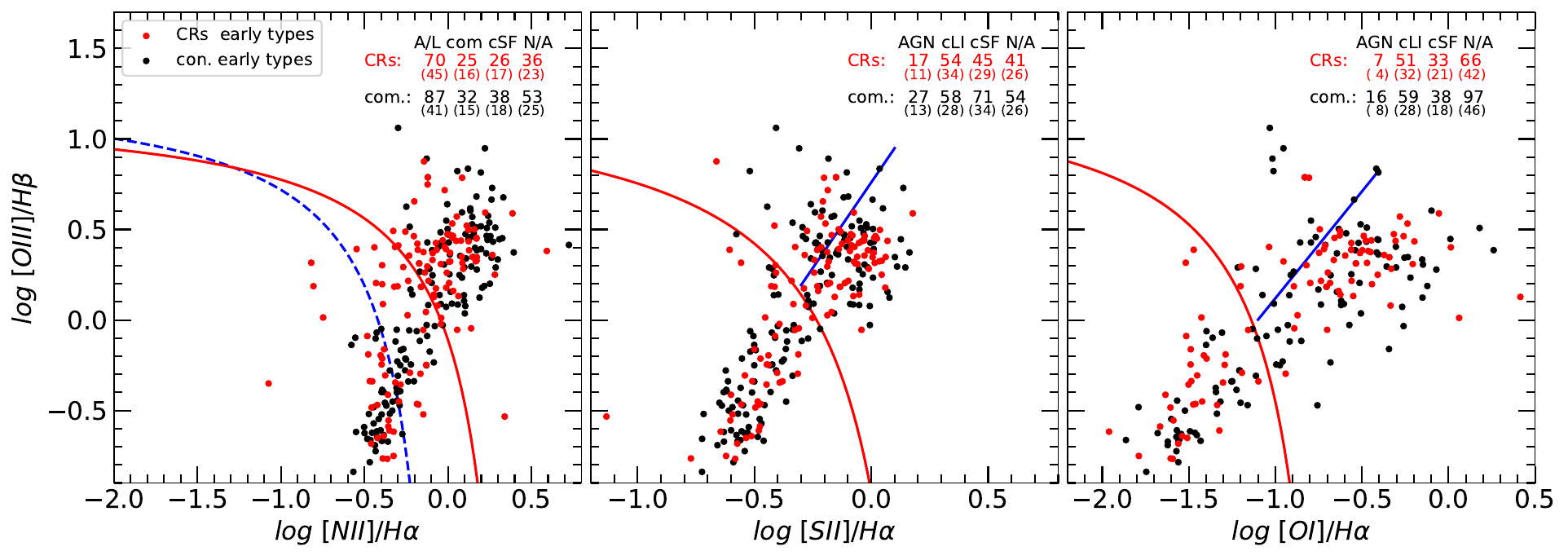}
\caption{Diagnostic diagrams plotting the emission line ratios
[NII]/H$\alpha$, [SII]/H$\alpha$, [OI]/H$\alpha$
and [OIII]/H$\beta$,  all for the central 2.5 arcsec region. 
The counter-rotating galaxies (red dots)
and the comparison sample (black dots)
are plotted with the conventional lines for 
classifications taken from \citet{BPTlines}. 
The number of galaxies in each class and their fraction 
as the percentile (in parenthesis)
are summarized in the upper-right corner of each panel. 
The fraction of cLIER (LINER) galaxies is higher in the counterrotators
than in the comparison sample.
\label{fig:BPT}}
\end{figure*}

The emission line ratios are analyzed with the standard diagnostic
diagrams, [NII]/H$\alpha$ vs.~[OIII]/H$\beta$ (the Baldwin–Phillips–Terlevich or BPT diagram), 
[SII]/H$\alpha$ vs.~[OIII]/H$\beta$ (the Veilleux-Osterbrock or VO diagram),
and [OI]/H$\alpha$ vs.~[OIII]/H$\beta$
\citep{BPTdiagram,VOdiagram}. 
Fig.~\ref{fig:BPT} shows the emission line ratios 
evaluated from the integrated spectra of the central 2.5 arcsec region. 
Only galaxies whose emission lines
are detected with a S/N greater than 3
are presented in Fig.~\ref{fig:BPT}
and hence each panel presents a somewhat different number of galaxies. 
The classification boundary lines on these diagrams are taken from \citet{BPTlines}. 
In this section, 
we analyze only the early-types of the counterrotators and the comparison sample 
to compare within the same morphological types, 
excluding S$_{unc}$ given the morphological uncertainty and Sp which barely exists in the counterrotators.

Fig.~\ref{fig:BPT} shows the diagnostic diagrams 
for the counterrotators and the comparison sample.
Based on the classification boundaries of \citet{BPTlines},
we divided galaxies into four groups.
The classification is summarised in Table \ref{tab:morp} with categories: active galactic nucleus (AGN), central low ionization emission regions(cLIER, also called LINER), star forming region in central area (cSF), and galaxies whose one (or more) major emission line is not detected (N/A).
The percentages of gaseous counter-rotating galaxies 
in each of these groups is 11, 34, 29 and  26\%, respectively, 
while for the comparison sample it is 13, 28, 34 and 26 \%.
The numbers counted for each classification are also presented 
in the upper-right corner of each panel.
Compared to the fractions of the comparison sample, the cLIER fraction 
in the counterrotators is larger, while the cSF fraction is smaller.
The fractions of AGN and N/A are similar in both samples.
Since the ionized gas emission from the centers is dominant in the counterrotators,
we have the result that a large fraction of the counterrotators have LIER-type characteristics even when the emission lines are measured from all IFU spaxels.
This agrees with the result of \citet{Ristea2022} 
that misalignment galaxies are more likely to have LIER emission lines.
We searched for high velocity dispersion 
in the major emission lines to test if shocks might be an ionizing source. 
Due to the spectral resolution and beam smearing effect of the MaNGA IFU observations, however, 
we could not constrain an ionizing source for the LIER emission at the center.

\subsection{The Star Formation Rate} \label{sec:SFR}

The star formation rate (SFR) is estimated from the $H\alpha$ luminosity 
using the calibration of \citet{Kennicutt1998,Kennicutt2009} with the method 
to estimate a \emph{pure} star formation rate of \citet{Blanc2009,Belfiore2016}.
This method estimates the fraction of the luminosity of $H\alpha$ 
emitted by the gas ionized only by the star formation activity, assuming that the emission line ratio of $[SII]/H\alpha$
represents the combination of gas ionized by star formation (SF) and LIER (L),
\begin{equation}
\left( \frac{[SII]}{H\alpha} \right) =  f_{SF} \left( \frac{[SII]}{H\alpha} \right)_{SF} + f_{L} \left( \frac{[SII]}{H\alpha} \right)_{L} 
\end{equation}
where $[SII]/H\alpha$ is the emission line ratio, and $f$ is the fraction of the $H\alpha$ luminosity ionized by each source (L for 'LIER' and SF for 'star formation'). 
The emission line ratios for each source are assumed as 
$( \frac{[SII]}{H\alpha} )_{SF}=0.4$ and 
$( \frac{[SII]}{H\alpha} )_{L}=1.0$. 
With these assumptions, the fraction of the $H\alpha$ luminosity stemming from star formation activity is given by
\begin{equation}
f_{SF} = (1-\left( \frac{[SII]}{H\alpha} \right))/0.6
\end{equation}
We applied this for each spaxel, and estimated the star formation rate 
from the corrected $H\alpha$ luminosity.
Note that the following characteristics about star formation rates
were also confirmed with the total $H\alpha$ luminosity without applying this method 
as well as the star formation rate estimated from the WISE W3 luminosity and
$H\alpha$ luminosity corrected by 22 $\micron$ luminosity using the
WISE W4 magnitude.
As in the previous section, we analyze early-types of the counterrotators and the comparison sample 
to compare within the same morphological types, 
excluding S$_{unc}$ and Sp.

\begin{figure}[ht!]
\centering
\includegraphics[scale=0.45]{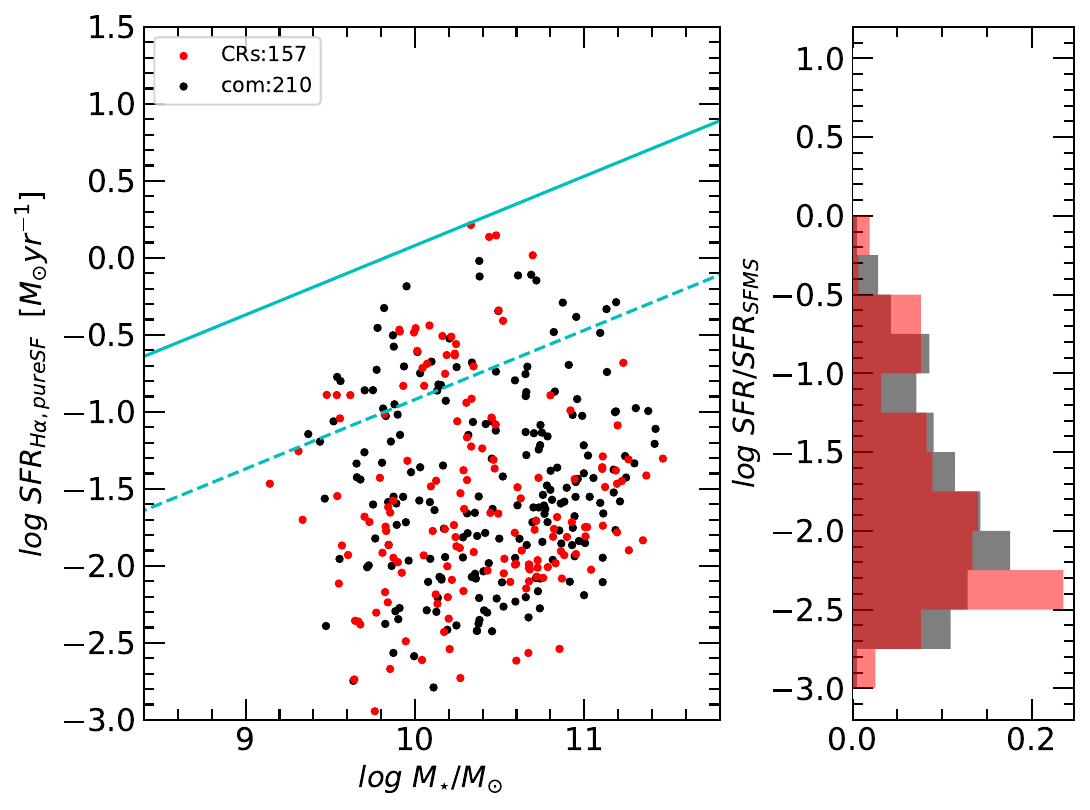}
\caption{The star formation rates (SFR) for the counterrotators and the comparison sample. 
The star forming main sequence taken from \citet{Davies2016} and its 10\% lines are shown as blue solid and dashed lines, respectively.
The right panel shows a histogram of the SFR compared to the SFR of the star forming main sequence line. 
The histogram has been normalized to show relative fractions 
because the comparison sample is about 1.5 times larger 
than the counterrotators.
\label{fig:SFR}}
\end{figure}

Fig.~\ref{fig:SFR} shows the star formation rate (SFR) with the stellar mass
and the histogram of the SFR compared to the SFR of
the star forming main sequence (SFMS). 
The SFR of the SFMS using the $H\alpha$ luminosity estimation calibrated by the RT method is taken from \citet{Davies2016}.
In the left panel, the SFRs for both the counterrotators and the comparison sample are similarly distributed on the diagram.
To see how weak the star formation is, 
we plot the SFR compared to SFR of the SFMS on the right histogram,
which also represents the deviation from the the SFMS line on the left diagram.
We can confirm the majority of the counterrotators and the comparison sample 
have weak star formation as shown as the largest peak is around -2.0.
And only small fraction of galaxies of both samples have 
a SFR close to the SFR of the SFMS.
When we define `suppressed star formation' galaxies as galaxies with the star formation rate of
10\% or lower than the star forming main sequence,
the fraction of the suppressed star formation galaxies 
is 82.2\% for the counterrotators (129/157) 
and 83.8\% for the comparison sample (176/210). 
As listed in the second-most right column of Table~\ref{tab:morp},
the SFR compared to the SFR of SFMS is
-1.89, -1.86 and -1.63 for E, E/S0, and S0 of the counterrotators, respectively.
These seem to be similar within the uncertainty, 
to -1.97, -1.73, and -1.62 for E, E/S0, and S0 of the comparison sample, respectively.
The characteristic of `weak star formation' appear to be
similar to each other \emph{on average}.

\begin{figure}[ht!]
\centering
\includegraphics[scale=0.45]{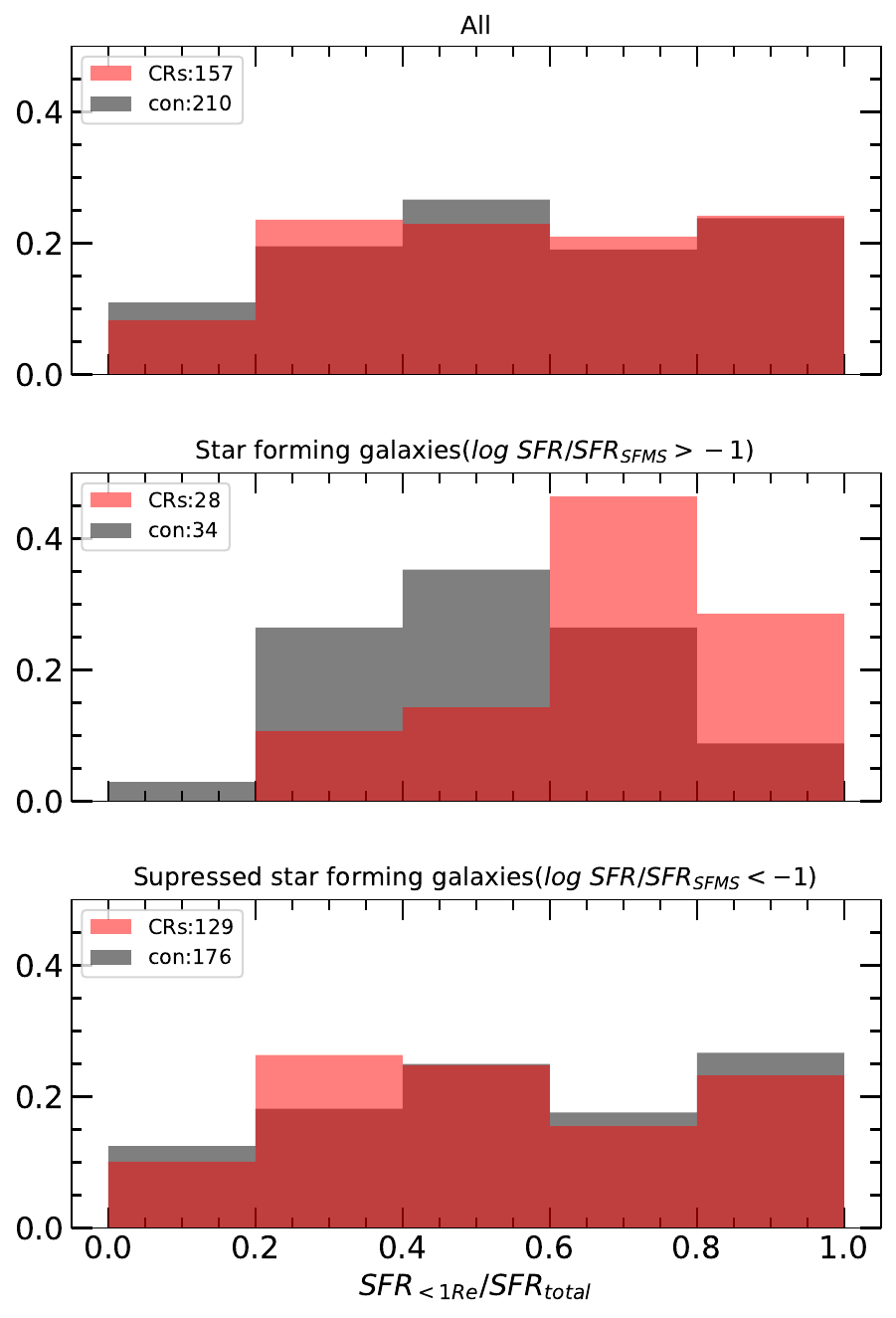}
\caption{The concentration of the star formation 
for groups with different star formation rates (SFR) 
with respect to the star forming main sequence (SFMS).
The concentration is measured by the fraction of the $H\alpha$ luminosity enclosed within 1 $R_{eff}$ to the total $H\alpha$ luminosity. 
The histogram of the counterrotators (red) and the comparison sample (gray) is constructed as three panels;
the top panel is for all galaxies in the analysis, and the middle panel
is for star forming galaxies whose total SFR is higher than 10\% of the SFR of SFMS. 
The bottom panel shows the histogram of galaxies with suppressed star formation.
All the histograms are presented as the relative fractions since the comparison sample is 1.5 times larger than the counterrotators (the number of each sample shown in the legend). 
The star forming counterrotators have high fraction of $SFR_{1Re}/SFR_{total}$, which indicates a high concentration of the star forming region in the center. 
\label{fig:SFR1Re}}
\end{figure}

We also analyze how the HII regions are concentrated at the center.
We evaluate the concentration of the star forming region at the center
through the fraction of the SFR within 1 $R_{eff}$ to its total SFR.
Similar to the previous analysis, we took 
SFR$_{H\alpha,pureSF}$ after the pure star formation correction.
Fig.~\ref{fig:SFR1Re} shows the histogram of the central concentration of 
the star forming regions of the counterrotators and the comparison sample.
In the top panel, 
the concentration of star forming regions is widely distributed 
in both the counterrotators and the comparison sample.
In the middle panel for `star forming galaxies' whose SFR is larger than
10\% of the SFR of the SFMS, however, the counterrotators have high central concentration values compared to those of the comparison sample.
In the bottom panel, which shows the suppressed star formation galaxies, 
no noticeable difference between the two samples
appears as in the top panel.
In summary, the star formation in the counterrotators  
seems to be more concentrated at the center if they have sufficient SFR.
This agrees well with the result of \citet{Chen2016} that 
blue gaseous counterrotators have strong central star formation, possibly
indicating angular momentum loss in the gas. 

In terms of morphology, 
we also compared the concentration of the star forming regions at the center
between the lenticulars and the ellipticals of each sample.
The star forming regions are more concentrated in the lenticulars
compared to that in the ellipticals for the counterrotators,
while the star forming regions are not as concentrated
in the lenticulars as well as the ellipticals of the comparison sample.
So, the lenticular counterrotators show the characteristic 
central concentration
of the star forming regions seen in the gaseous counterrotators.

\subsection{Stellar and Gas Phase Metallicity} \label{sec:metallicity}

We compared the stellar metallicity of the counterrotators and the comparison sample
using $[MgbFe]'$ and $Mgb/\langle Fe \rangle$ taken from  \citet{Burstein1984,Thomas2003}.
Both samples have similar distributions on the metallicity ($[MgbFe]'$) and
$\alpha / \langle Fe \rangle$ ($Mgb/\langle Fe \rangle$) without any evidence that 
one sample is metal-poor than the other.
Also, their radial gradient of the stellar metallicity in each galaxy is similar in both the sample.
Therefore we conclude that the stellar metallicity of the gaseous counterrotators is just following that of typical early-type galaxies.


We also compared the gas phase metallicity of the counterrotators and the comparison sample. 
Similar to the stellar metallicity, there is no significant difference in the gas phase metallicity between the two samples. 
This agrees with the results of \citet{Beom2022,Zinchenko2023}.
However, we would like to emphasize that this does not preclude
possible pristine gas accretion as the origin of the counter-rotating gas.
In the counterrotators, the ionized gas is generally concentrated at the center
so that the gas phase metallicity is representative of 
the central gas where the gas has been mixed up with the pre-existing gas 
and enriched through star formation and gas recycling.
To search for evidence of pristine gas accretion,
one would need to analyze the gas phase metallicity at the outskirt
of the counterrotators with sufficient sensitivity to compensate
for the weak ionized gas emission.

\subsection{Star Formation Quenching and a Central Poststarburst Signature} \label{sec:qunching_PSBR}

We used Pipe3D spectral fits to analyze stellar populations in the MaNGA spectra 
for inner and outer regions in the galaxies.
We not only compared the mean stellar age, but also 
calculated the cumulative age distribution of the stellar populations inferred for each region.
All the analyses show no significant difference between the stellar populations 
of the counterrotators and the comparison sample at the inner region (within 1 $R_{eff}$) as well as the outer region.

We searched for a poststarburst spectral signature using the definition of 
EW(H$\delta$) $>$ 3~\AA~ in stellar absorption, 
EW(H$\alpha$) $<$ 10~\AA~ in emission,
and $\rm log~EW(H\alpha) < 0.23 \times EW(H\delta) - 0.46$ \citep{Chen2019}.
Note that the equivalent widths of both the H$\delta$ {\it absorption} 
and the H$\alpha$ {\it emission} lines
are defined to be positive with these criteria. 
A poststarburst signature at the center is found 
in five counterrotators and only one from the comparison sample. 
Although the number of post starburst galaxies is small, 
we can conclude that a central post starburst signature seems
more common for the counterrotators, also
because the sample size of the comparison sample is
about 1.86 times bigger than that of the counterrotators 
(about 1.34 times bigger if considering only early-types).

\subsection{Environment} \label{sec:groups}

We analyze the environment of the counter-rotating galaxies 
by identifying group members of the sample.
We identified the group members by the projected distance on the sky
and the velocity difference based on the spectroscopic redshift 
taken from SDSS fiber spectroscopy of DR12 (\citet{SDSSDR12}) and MaNGA IFU observations.
Applying the friend-of-friend (FoF) method taken from \citet{Tempel2017}, 
we counted group members of the sample.
Note that the number of group members that might miss galaxies whose spectroscopic redshift is not observed in SDSS and MaNGA. 
Also, almost all of the MaNGA galaxies are in a group environment.
We did a cross-check using the galaxy group catalog of \citet{Tempel2017}
and applied a fixed box method (500~kpc and 500~km/s). These produced 
similar results to those described in this section. 
We analyzed the number of members in a galaxy group hosting 
the counterrotators, and did the same for the comparison sample.
We found no noticeable difference in the number of group members between 
the counterrotators and the comparison sample.
The only difference is that the counterrotators are somewhat more frequently found as isolated or with only a single group member; however,
this result was not significant based on a KS test.
This agrees well with the result of the previous study of \citet{Bettoni2001} that no significant differences in the environment appear
between the counterrotators and the control sample.

We also checked the compactness of the group environment 
(the standard deviation of the projected distances and 
the relative velocity differences from the group center), 
which shows no noticeable difference 
between the groups hosting the counterrotators and the control sample.
The distance to the nearest group member from each counterrotator 
is also not significantly different from that estimated of the comparison sample.
Thus, we would conclude there is no evidence that a counterrotator is more likely 
to be formed in more compact group environment.

\section{Unique Gaseous Counterrotators} \label{sec:Unique}

In this section, we discuss a few details for some interesting 
gaseous counterrotators that may provide important clues on their formation scenarios.

\begin{figure*}[ht!]
\centering
\includegraphics[scale=0.47]{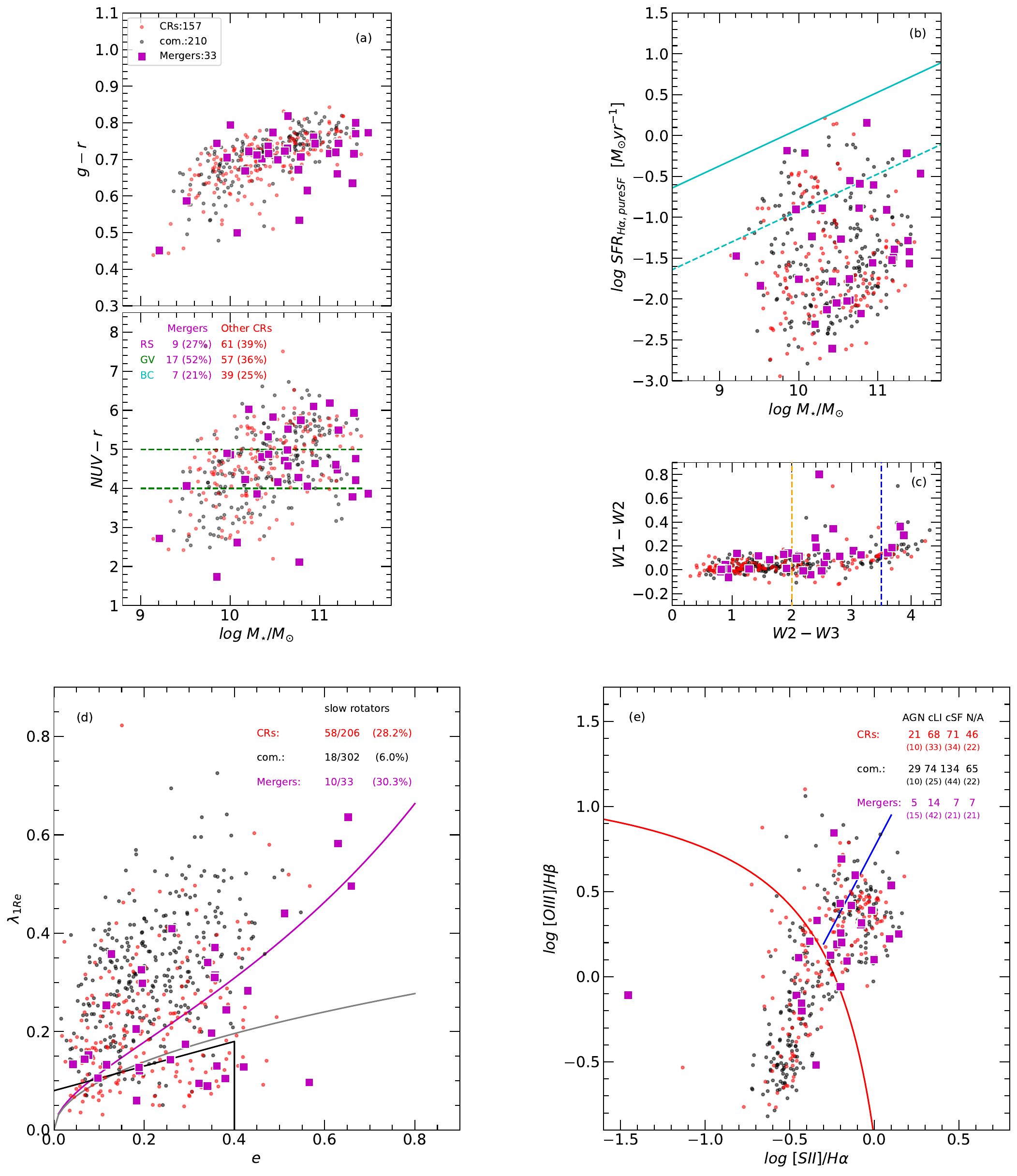}
\caption{ Physical properties of merger candidates among the counterrotators compared to the other counterrotators and the comparison sample (which was selected not to have potential merger signatures).
Panels (a) through (e) use the same parameters and configuration of the diagrams as Fig.~\ref{fig:CMD_morp}, \ref{fig:SFR}, \ref{fig:wise}, \ref{fig:spin}, and \ref{fig:BPT}.
Merger candidates are presented as purple squares, and 
other counterrotators and comparison samples are 
shown as transparent red and black dots, respectively. 
The numbers in the legend on the panel (a) represent the number in each sample.
\label{fig:mergers}}
\end{figure*}

\subsection{Mergers Candidates} \label{sec:Mergers}

Since a merger with a gas-rich galaxy may be a promising mechanism to 
produce counter-rotating gas disks, we searched for signatures indicative of ongoing or recent interaction. 
The fraction of merger candidates among counterrotators 
may constrain how common a merger origin is for their formation,
as well as provide information on the formation time scale for a settled counter-rotating disk.
We identified merger candidates whose stellar or ionized gas structure indicates recent or ongoing merging interaction as follows:
1) a stellar or ionized gas stream between a counterrotator and a nearby companion galaxy; 
2) a companion galaxy located within about 100 kpc in projected distance on the sky and its receding velocity aligned with the rotating direction of gas kinematics in the counterrotator;
or 3) a significantly asymmetric shape or tidal stream
in the stellar light distribution on broadband images,
if no nearby companion galaxy was present.
With these definitions, we find 33 merger candidates, which 
corresponds to only about 11\% of all gaseous counterrotators.
However, the ability to detect a disturbed light distribution or bridges of material between the counterrotators and nearby companions is quite limited with the existing data, and undoubtedly the actual fraction of merger candidates will be higher. Much deeper imaging and HI data, for example, would be useful to detect more merger candidates.
Among our `merger candidates' there are two examples of a counterrotator
with two companions located close enough that it could be a triple galaxy interaction,
and three cases of a counterrotator located in a compact group 
with more than three galaxies.

The merger candidates were analyzed 
using the same methods described in the previous sections, and their properties are shown in the multiple panels in Fig.~\ref{fig:mergers}.
As described in the previous sections,
the majority of the counterrotators are early-type galaxies
and the general characteristics resemble the physical properties of early-type galaxies.
Therefore, we compare the physical properties of the merger candidates to
those of the early-type counterrotators and the early-type comparison sample.
Almost all the merger candidates 
do not have evidence of spiral arms.
Only one merger candidate has spiral arms and we will discuss it
in Section \ref{sec:spirals} in detail.
In panel (a) of Fig.~\ref{fig:mergers} 51\% (17/33) of the mergers are 
in the green valley regime in the NUV-r color versus stellar mass diagram, 
a higher fraction than 
36\% (57/157) for all early-type counterrotators. 
We did not find a trend in the mergers
for a more prominent young stellar component, which would appear as a blue color (in panel (a)) or a higher current star formation (panel (b)).
We conclude that star formation is not enhanced in the merger candidate counterrotators. 
More of the merger candidates tend to have intermediate dust contents where the W2-W3 color is between 2 and 3,
while the other counterrotators are located more in the poor dust regime on the diagram where W2-W3 color is less than 2.
However, due to the small sample size of the merger candidates 
it is hard to conclude that the merger candidates 
tend to be more dust-rich than the other counterrotators.
We found relatively low spin parameter values for the mergers 
and a high fraction of the slow rotators, similar as for the other counterrotators.
If we consider assume that the merger candidates may show a younger stage after the formation of the gas counter-rotation, 
the characteristic of `low spin parameter' was already induced
in an early time during the formation and had since on average stayed the same, not evolving slowly or decreasing. 
A characteristic of the mergers that is distinct from the other counterrotators 
is a high fraction of AGN galaxies (15\%, 5 out of 33) and cLIER/LINER galaxies  (42\%, 14 out of 33).
The combined AGN and cLIER category then is 57.6\% (19/33), higher than 43.2\% for the total sample of counterrotators.
The high AGN and LINER/cLIER fractions may support the idea that
gaseous counterrotators can trigger AGN feeding through concentration of gas at the center through angular momentum loss.
However, the difference is only marginally significant, given the small number of the merger candidates.
In addition, as we described in Section \ref{sec:BPT}, 
the origin of the ionized gas
could indicate shocks or diffuse Ionized gas.

\subsection{Stellar Counterrotators} \label{sec:SCR}

As introduced in section \ref{sec:identification}, 
the gaseous counter-rotation can have two stellar components kinematically decoupled,
the cases b-i) and b-ii) in Fig.~\ref{fig:CRs_sample}.
One stellar component is rotating in the opposite 
direction to the other stellar component, hence co-rotating with the gas.
If the stellar component co-rotating with the gas is located in the inner part
(as shown in b-i)), we will call it stellar counterrotators with inner co-rotation.
For the other case (as shown in b-ii)), with the co-rotation at the outer part,
we will call it the stellar counterrotators with outer co-rotation.
The former cases are confirmed in 25 gaseous counterrotators,
and the latter cases are present in 4 gaseous counterrotators.
Note that we might miss some cases due to the IFU coverage, 
especially in the `primary sample' of MaNGA (observed up to 1.5 R$_{eff}$).
In this section, we will focus more on the former case (the inner co-rotation case)
not only because they are about 6 times more frequent,
but also because they
provide interesting clues to understand the formation of the gaseous counterrotators.

\begin{figure*}[ht!]
\centering
\includegraphics[scale=0.50]{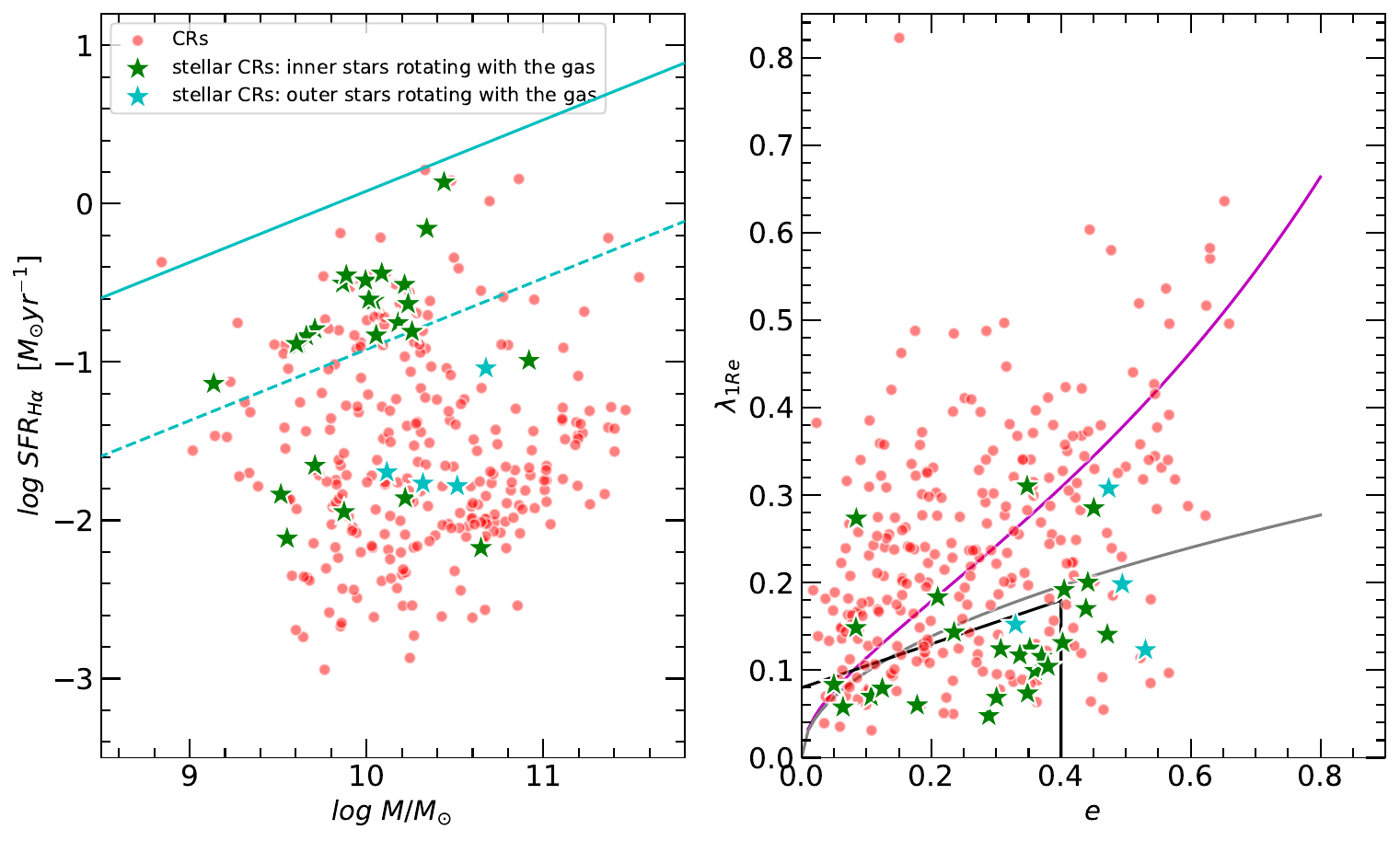}
\caption{The star formation rate and the spin parameter 
for galaxies that are a stellar counterrotator as well as a gaseous counterrotator. 
These systems have a minor stellar component counter-rotating to the major stellar component.
The minor component is co-rotating with the gas component 
near the center (green stars) or 
in the outer region (light blue stars).
The red dots are all the other gaseous counterrotators.
The left and right panels are made
as in Fig.~\ref{fig:SFR} and \ref{fig:spin}, respectively.
\label{fig:SCR}}
\end{figure*}

Fig.~\ref{fig:SCR} shows the star formation rate (SFR) and the spin parameter
for the stellar counterrotators, the other early-type  counterrotators and the comparison sample.
Among 25 stellar counterrotators with inner co-rotation, 17 cases show relatively high star formation rates,
larger than 10\% of the SFR of the star forming main sequence (above the dashed line in the left panel of Fig.~\ref{fig:SCR}).
This corresponds to 68\%, which is significantly 
higher (about 3 times) than that for the other counterrotators.
Their star formation regions are also concentrated at the center. 
In addition, 19 galaxies out of 25 are classified as slow rotators.
These are located under the gray line in the right panel of Fig.~\ref{fig:SCR}.
This corresponds to 76\%, about 2.5 times larger than that for the other counterrotators.
Thus stellar counterrotators with inner co-rotation
show distinct characteristics from those of other
gaseous counterrotators.
They may be cases of
gaseous counterrotators where pre-existing gas was abundant to induce the gas collision and gas concentration at the center, fueling star formation.

The minor stellar component at the center kinematically coupled with the gas
suggests there was significant star formation in the past
as a result of the accretion of the counter-rotating gas.
A stellar population synthesis that includes two kinematically decoupled
populations can estimate when the counter-rotating gas was accreted 
and how much star formation resulted in this specific process. 
This is beyond the scope of this paper, but
we tried a rough estimate.
We assume the star forming targets, 17 green star symbols above the dashed line on the left panel of Fig.~\ref{fig:SCR}, are in the early stage after the formation,
while the targets with the suppressed star formation, 8 green star symbols 
under the dashed line on the panel, are in
the later stage with much reduced star formation. 
Using the stellar population synthesis information of the pipe3D, \citet{pipe3D}, 
we compare their central stellar population within 1 $R_{eff}$.
As their current star formation rate implies, 
the 17 galaxies in the early stage have young stellar population by 25\% 
formed as recently as the past several million years.
In contrast, the 8 targets with the suppressed star formation have 
20\% of their central stellar population formed within the past 1 Gyr.
From this rough estimate, we may deduce that the star formation phase is short lived 
and that quenching seems to happen within about one Gyr
for these cases of stellar counterrotators with inner co-rotation.

\subsection{Spiral Counterrotators with the Grand Design of Spiral Arms} \label{sec:spirals}

\begin{figure*}[ht!]
\centering
\gridline{
          \fig{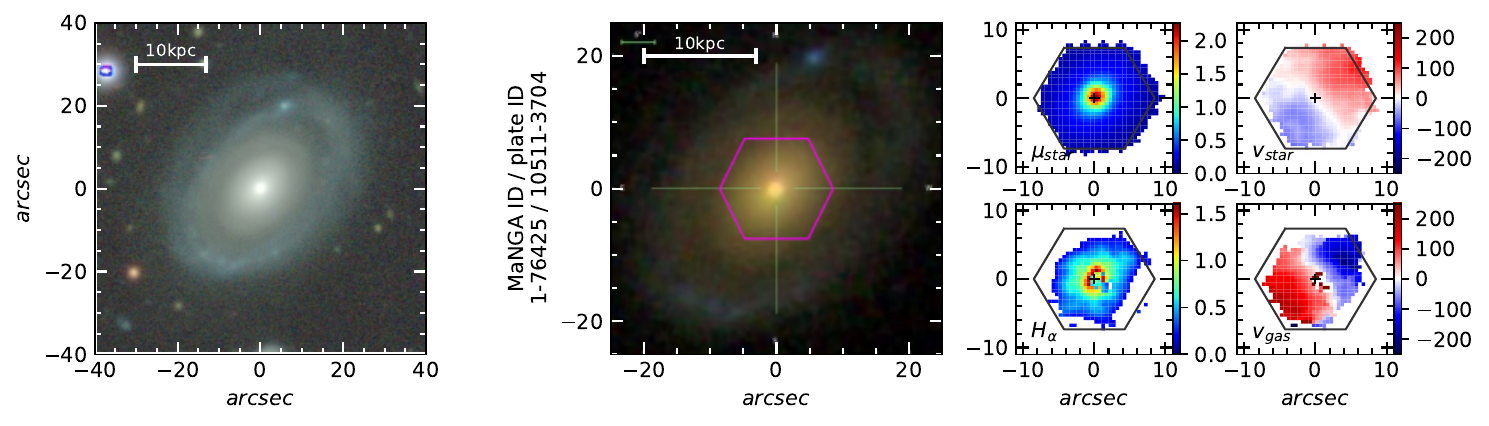}{0.90\textwidth}{(a) 
          A spiral galaxy with the gaseous counter-rotation}
          }
\gridline{
          \fig{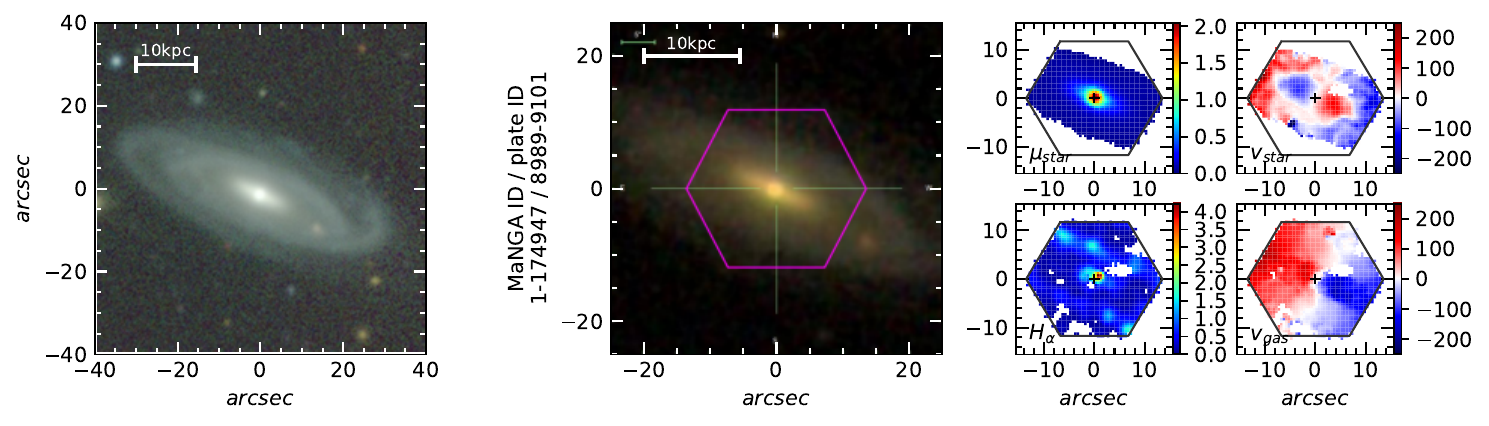}{0.90\textwidth}{(b) 
          The spiral galaxy with the stellar counter-rotation as well as the gaseous counter-rotation }
          }
\gridline{
          \fig{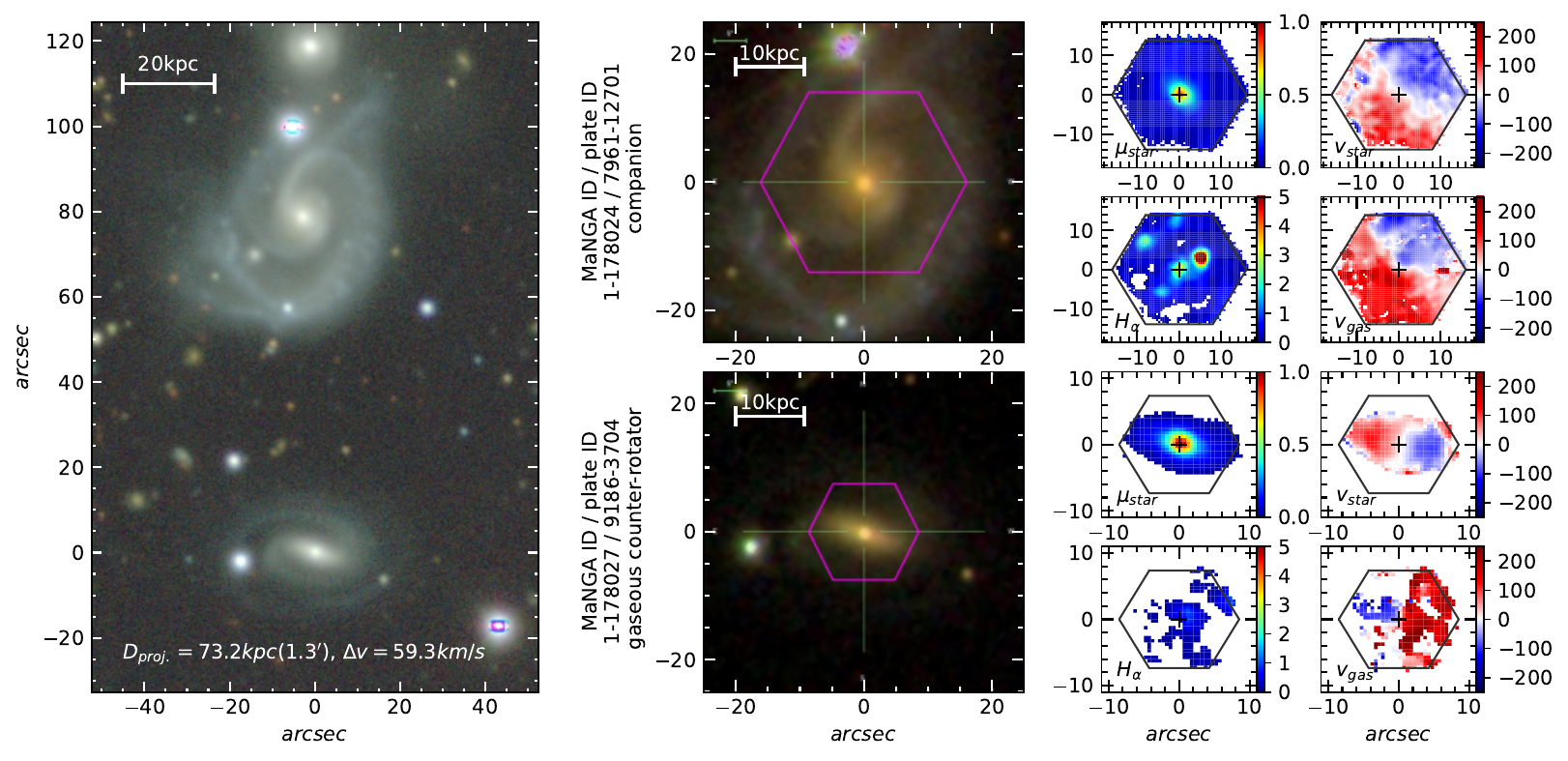}{0.90\textwidth}{(c) 
          The spiral galaxy as a merger with its companion as well as the gaseous counter-rotation}
          }
\caption{The three spiral galaxies with a gaseous counter-rotating component. 
For each counterrotator, DECaLS and SDSS images are the left panels.
The third panel of figures shows the surface brightness in the g-band and in H$\alpha$ emission, in units of 10$^{-17}$ $erg$ $s^{-1}$ $spaxel^{-1}$ $\AA^{-1}$ and 
10$^{-17}$ $erg$ $s^{-1}$ $spaxel^{-1}$, respectively.
The rightmost panels show the velocity maps of the stellar and 
gaseous components in units of $km$ $s^{-1}$.
The panels in (c) are arranged in almost the same configuration as (a) and (b), 
but a DECaLS image of a larger field of view shows the two galaxies 
in the merging system.
\label{fig:spirals}}
\end{figure*}

In this section, we introduce three unique spiral galaxies among the gaseous counterrotators.
Their spiral arms show a grand design of typical spirals,
and this is infrequent among the gaseous counterrotators.
Two spirals show no nearby companion or
feature suspicious of recent or ongoing mergers,
while the other spiral is in a merging phase with its companion. 
Fig.~\ref{fig:spirals} shows the three spirals 
with images from DECaLS and SDSS, and stellar and gas components from MaNGA.

The first spiral shown on panel (a) in Fig.~\ref{fig:spirals}
resembles a lenticular with ring structures 
in the outer disk. This is not rare for counterrotators as mentioned 
in Section \ref{sec:Sample}. 
However, the structure linking the central region to the outer ring
and the appearance of the outer ring look like spiral arms.
Its central stellar component is similar to that of a typical lenticular;
the bulge-to-total light ratio is 0.47, with a smooth light profile for the central disk, but its S\'ersic index for the bulge is somewhat low as 2.8.
A central region of a lenticular and outer spiral arm structure may imply this galaxy could be in the transition phase between a spiral and a lenticular.
Since only three spiral galaxies are identified among 303 of the counterrotators,
the formation of spiral arms after the formation of the counter-rotating gas seems not common and hence not a likely explanation for this object.
This target is located on the red sequence (g-r color is about 0.7)
with relatively higher stellar mass compared to the other counterrotators 
($log$ $M/M_{\odot}=10.60$).
It is also dust-poor based on the W2-W4 color shown in the fourth column of Fig.~\ref{fig:wise}.
It also has low HI mass with no detection by HI-MaNGA
(the upper limit of HI appears to $log~M_{HI,lim}=9.44$) \citep{Stark2021}.
Its spin parameter is not high but also not low enough to consider it a slow rotator.
The major emission line strengths are too weak to be classified using the diagnostic diagram. 
The star formation is suppressed star formation at 0.0129 $M_{\odot}~yr^{-1}$ ($log$ $SFR/SFR_{SFMS}$ = -2.2).
However, this is only from the central region covered by the IFU observation
(purple hexagon in the second panel from the left of the SDSS image of Fig.~\ref{fig:spirals}).
Considering the stellar and dust structure of the outer spiral arm or ring on the DECaLS image on the leftmost panel,
significant star formation could exist in the outer arm and ring.
We also note that, since the IFU covers only the central region,
this counterrotator may have different kinematics 
in the outer disk where spiral arms reside, such as stellar or gaseous counter-rotation
with respect to those of central region,
as the following second spiral shows.

The second spiral is presented in panel (b) in Fig.~\ref{fig:spirals}.
It shows clear spiral arms in the disk with a smooth central light profile. 
The overall characteristics are quite similar to the first spiral described above.
The bulge-to-total light ratio is 0.5, and the S\'ersic index of the bulge is 5.7,
comparable to those of a typical lenticular.
This target is also on the red sequence with relatively high mass 
with g-r color of 0.7 and log stellar mass of 10.67.
It appears as a dust-poor galaxy based on W2-W3 color (the fourth column of Fig.~\ref{fig:wise}), and not a slow rotator.
Its HI is detected by MaNGA-HI as 1.07 $\times$ 10$^{10}$ $M_{\odot}$,
which corresponds to the ratio of HI mass to stellar mass of 0.46.
The star formation rate is 0.0912 $M_{\odot}~yr^{-1}$ ($log$ $SFR/SFR_{SFMS}$ = -1.4),
which also does not include the star formation rate in some parts of 
the spiral arms due to the IFU coverage.
The most noticeable feature of this target is 
the stellar counter-rotating component in the center,
which is introduced in the section \ref{sec:Sample} as (b-ii).
This kind of stellar counter-rotation is confirmed only in 4 cases 
among the gaseous counterrotators. 
Since the stellar rotation co-rotating with the gas is located in the outer part,
this galaxy seems to have a formation scenario totally different from that described in Section \ref{sec:SCR}. It seems a good candidate for a dry merger with a stellar companion that only produced a central stellar counter-rotating component.

The other spiral galaxy is in a merging phase with its nearby companion.
As seen in the DECaLS image in Fig.~\ref{fig:spirals},
the gaseous counterrotator with two spiral arms 
is located on the bottom, with its nearby companion
located at a projected distance of 73.2 $kpc$ 
and a velocity difference of 59.3 $km~s^{-1}$.
The two galaxies are close to face-on, so that
it is difficult to know the merging direction with respect to 
the direction of the gas rotation in the gaseous counterrotator.
Based on significantly disturbed features on the spiral arms in the companion, 
two galaxies are in an ongoing galaxy interaction.
Another galaxy is located on the top edge of the DECaLS image.
We do not have a spectroscopic redshift and
there is a possibility this merging system consists of three galaxies.
Since the pre-existing gas has to be swiped by the gas acquired to form a gaseous counterrotator,
a spiral galaxy is not a good candidate for a host to be a gaseous counterrotator.
This case, however, shows that a spiral galaxy can be the host 
to form a gaseous counterrotator.
The companion is 5 times larger than the spiral gaseous counterrotator 
in terms of stellar mass ($log$ $M/M_{\odot}$ is 11.13 to 10.43). 
The g-r color of the counterrotator is 0.72, redder than 0.65 of the companion.
Both galaxies were observed by HI-MaNGA using GBT (Green Bank Telescope).
Since the beam of the GBT covers these galaxies and these pair are located 
close each other, 
the HI masses measured is the total HI mass of the pair galaxies.
The HI mass of the system was observed twice 
by 1.80 $\times$ 10$^{10} M_{\odot}$ pointing to the counterrotator 
and 2.38 $\times$ 10$^{10}M_{\odot}$ pointing to the companion. 
The properties of the companion, a gas-rich massive spiral galaxy, 
might imply that the galaxy showing gaseous counter-rotation may not survive if a full merger happens. This system is an interesting target for deep, high resolution HI observations to better characterize the interaction between the two objects.

\section{Discussion} \label{sec:Discussion}

\subsection{The Frequency and Morphology of Gaseous Counterrotators} \label{sec:discussion_morp}

Among 9992 MaNGA galaxy targets of the final version of MaNGA data, Data Release 17 (MPL-11), 
we identified 303 gaseous counterrotators whose 
misalignment angle between stars and gas is larger than 150 degrees.
This corresponds to 3\%.
Based on the previous IFU studies on all misalignment galaxies, 
(the misalignment angle between gas and stars is larger than 30 degrees),
the fraction of gaseous counter-rotating galaxies is reported as about 1.7\% (SAMI;\citet{Bryant2019,Ristea2022}) and about 2\% (MaNGA earlier versions;\citet{Li2021,Xu2022,Zhou2022}) from among the total number of galaxies. 
These are significantly lower than our value of 3\%.
This might be caused to some extent by the estimation of the kinematic position angle, which can become uncertain for less inclined galaxies or galaxies with irregular gas distributions and low S/N observations.
Estimating the fractions using only galaxies whose kinematic position angle could be determined through a fit,
the fraction becomes larger, up to about 3.5\% for SAMI as found by \citet{Bryant2019,Ristea2022}, and up to
3.0\% from the earlier MaNGA  data releases of \citet{Li2021,Xu2022,Zhou2022}.
Applying this criterion to our sample,
the fraction of gaseous counterrotators is about 3.8\%, reducing the discrepancy with previous results, though the difference is still larger than
the statistical uncertainty.

We classified the morphology of the counterrotators and the comparison samples
using results from \citet{Vazquez-Mata2022}, in order to
compare the physical properties within the same morphological class.
The majority (about 80\%) of the counterrotators are early-type galaxies.
The high fraction of gaseous counterrotators in early-types agrees with the previous statistics and the interpretation 
that the gas component in early-types is likely to have an outside origin 
\citep{KannappanFabricant2001,Krajnovic2011,Cappellari2016,Bryant2019,Ristea2022}. 
As mentioned in section 3.3.3, 
we found that gaseous counterrotators consist of about 5.58\% of the early-type galaxies in MaNGA, 
and about 12.6\% if we count only early-types with overall rotation in both gas and stellar components. 
This 12.6\% is somewhat higher than the previous results using SAMI data,
9.6\% (9/107) in \citet{Bryant2019} and 8.4\% (38/452) in \citet{Ristea2022}.
The ratio of ellipticals (E and E/S0) to lenticulars (S0) is
1.2 in the gaseous counterrotators.
Since lenticular galaxies are relatively rare (about 1/3) compared to ellipticals 
in the total galaxy samples in MaNGA, 
this comparable number of the lenticular counterrotators 
might tell us that gaseous counterrotators favor lenticular galaxies
over elliptical galaxies.

The fact that the majority of the gaseous counterrotators are early-types
implies their formation 
is highly correlated with the evolution of early-type galaxies.
There are two possibilities:
1) Counterrotators formed in an early-type host, or 2) turning a late-type galaxy into an early-type as a result of the gas accretion.
The former is evident. A gas poor galaxy is a good candidate host 
where a gas component with opposite angular momentum to the host's spin 
can be accreted and survive with the opposite angular momentum even after a collision
with any limited pre-existing gas of the host galaxy.
For the second case, the formation process of the gaseous counterrotator would have to lead to the destruction of spiral arms and remove pre-existing gas, and then turn into a largely quenched early-type galaxy with low remaining SFR. 
In addition, 
a gaseous counterrotator seems to barely produce (or take a long time to produce) any form of spiral arms in the gas and stars, 
regardless if spiral arms were destroyed or did not exist before the formation.

We note that well-structured spiral arms are very rare, so that only several spiral counterrotators have them.
Also, there is only one gaseous counterrotator in late spirals (T-type$>$3).
These might imply that the formation process of gaseous counterrotators destroys the spiral arms if it was in the ancestor galaxy and produces the environmental condition of the counter-rotating gas where spiral arms can not be formed.
The spirals, the minority of the gaseous counterrotators 
will be discussed in a future paper in more detail.

\subsection{Morphological Characteristics of the Gaseous Counterrotators} \label{sec:discussion_Morp_char}

The gaseous counterrotators are mostly identified as early-types, and their galactic properties resemble those of early-types. 
This is also seen in previous studies of misalignment galaxies, whose gas is misaligned by more than 30 degrees.
For example, considering their mass, misalignment galaxies are known to be relatively small, have a high concentration of the stellar component (high S\'ersic index and B/T ratio), low dust content and low HI detection rates, and a high fraction of LINER/LIER galaxies, weak SFR, and low $v/\sigma$ (low spin parameter)
\citep{Xu2022,Zhou2022,Ristea2022}.
Most of these properties are also confirmed in the edge-on counterrotators \citep{Beom2022}.
In this study, we attempted to disentangle
morphological effects from the effects of the counter-rotating gas.
Hence, we carefully classified the morphology of the gaseous counterrotators and of the comparison sample to allow comparison of their physical properties within the same morphological class.
Our results show that no significant difference between the corotators and the counterrotators is found for the size and concentration of the stellar component, for the dust content (low), and for the detection rate of HI (also low).
This suggests that those characteristics are primarily morphological effects, not necessarily a result of gas acquisition in the opposite direction of the galaxy's spin.
We also find that the gaseous counterrotators have three characteristics distinct from the corotators (comparison sample); 
weak galaxy rotation (low spin parameter for the stellar component),
high concentration of star forming regions at the center, if present,
and higher LINER/cLIER galaxy fraction.
These differences are from a comparison within the same morphological class,
so they likely arose from the impact of the acquisition of counter-rotating gas.
We will discuss them one by one 
in the following sections \ref{sec:discussion_spin_parameter}, \ref{sec:discussion_centralSF}, and \ref{sec:discussion_cLIER}.

The fact that there are several similarities (structure, low dust content, low HI content) between the counterrotators
and the corotators (the comparison sample) for the early-type galaxies (ellipticals and lenticulars) might tell us that for both samples, the gas 
likely had outside origins, with accretion onto relatively gas-free early-types. 
After all, there is no reason to expect external gas accretion to be limited to counter-rotation only, and it is likely that all early-type galaxies acquired their gasesous rotating disks in a later phase.  However, the differences between the two samples also show that the consequences of this acquisition of gas led to statistically different properties for some other parameters.

\begin{figure*}[ht!]
\centering
\includegraphics[scale=0.50]{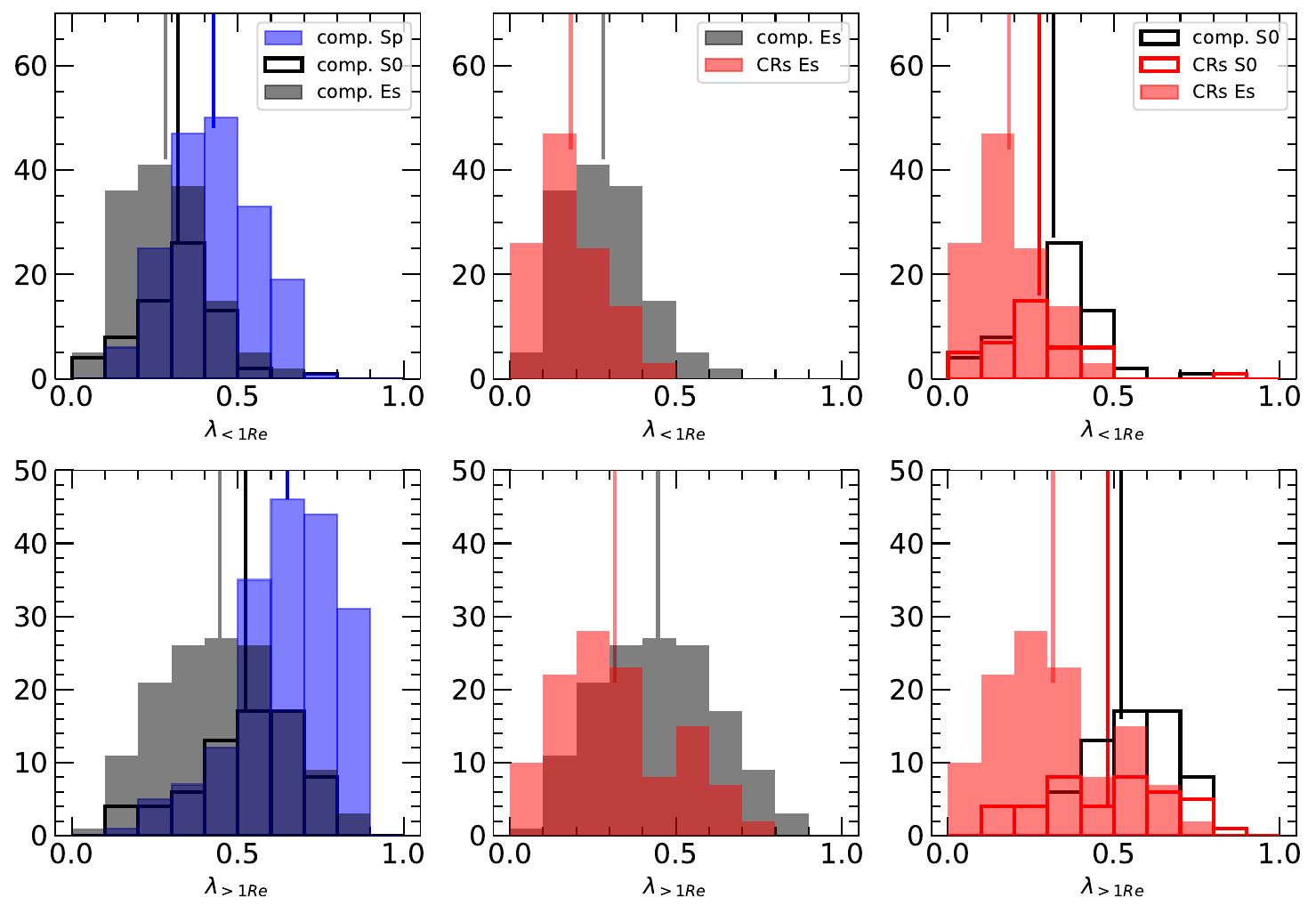}
\caption{The histograms of the spin parameters estimated in the inner region and out region depending on the sample and the morphology.
The upper panels and the bottom panels are for inner region (within 1 $R_{eff}$) and outskirt (between 1 and 1.5 $R_{eff}$).
The left panels show the distribution of the spin parameters for
corotators (the comparison sample) depending on the morphologies; 
spirals (blue), S0 (black box), and ellipticals (gray area).
The center panel shows the spin parameters for the ellipticals
in the corotators (gray area) and counterrotators (red area).
The right panel shows the spin parameters for 
S0 counterrotators (red box), S0 corotators (black box),
and elliptical counterrotators.
The vertical lines show the mean value of each distribution.
\label{fig:spin_hist}}
\end{figure*}

\subsection{Low Spin Parameter Values of Gaseous Counterrotators and the Merger Origin} \label{sec:discussion_spin_parameter}

The most significant characteristic of
the gaseous counterrotators is the tendency for lower spin parameters for their stellar components.
In this section, we compare the spin parameter values 
using histograms and discuss how they can be understood in the merger origin.
We use histograms, 
not the spin parameter versus the ellpiticity plane because 
the spin parameter depends on the ellipticity or orientation (inclination; for the details, see \citet{Cappellari2007,Cappellari2016}).
For example, a spin parameter value of 0.2 is differently interpreted 
for a galaxy with an ellipticity of 0.1 versus another galaxy with an ellipticity of 0.5. 
We therefore show the comparison of the spin parameter 
using histograms only to see if there are differences between the two samples. The two samples should have similar inclination distributions since inclination was not a selection criterion.
In addition, we verified
we get the same conclusion if we exclude low ellipticity galaxies where the orientation effect is maximal.

Fig.~\ref{fig:spin_hist} shows the spin parameter values measured 
for the inner region (within 1 $R_{eff}$) and for the outer region (1 to 1.5 $R_{eff}$)
depending on the sample sets and their morphology.
The left panels show the morphological effect on 
the spin parameters of the comparison sample.
As seen in their distributions, 
the spin parameter appears to be larger in the order of 
ellipticals, lenticulars, and spirals,
for both inner (upper-left panel) and outer regions (left bottom panel).

The center panels show the spin parameters for ellipticals 
for counterrotators and co-rotation (comparison sample).
In the central upper panel, 
elliptical counterrotators have smaller spin parameters in the inner region
than elliptical comparison galaxies.
This difference appears larger than that stemming from
the morphological effect between ellipticals and lenticulars 
in the comparison sample on the upper-left panel.
We also confirm that the elliptical counterrotators have 
small spin parameters even in the outer region compared to those of the corotators.
As addressed in Section \ref{sec:SFR}, 
a new stellar population can be formed in the gas component driven into the center.
Since the new stellar population has counter-rotation with respect to the spin of the other stellar population, the stellar absorption lines are broadened.
If this is a dominant cause, 
the spin parameter would be significantly small only for the inner region 
where the new star formation was active.
However, the counterrotators have low spin parameters for the outer region 
as well as the inner region compared to those of the corotators.
The discrepancy between the two samples is even larger in the outer region 
(the central bottom panel) than in the inner region (the central upper panel).

Another mechanism to induce low spin parameter value is a
merger origin, which can cause a change in the kinematics over the whole galaxy.
Early galaxy interaction studies reported that 
the ratio of rotational velocity to velocity dispersion 
in the stellar component
could be changed by the merging conditions (mass ratio, galaxy spin directions with respect to each other, inclination, gas content, etc.;
\citet{Bournaud2005,Cox2006}). 
In particular, a low ratio between the rotational velocity to the velocity dispersion ($V/\sigma$)
can be induced only by major mergers rather than minor mergers.
Therefore, the low spin parameters might indicate 
that a major merger is the most probable formation origin,
in particular for the elliptical gaseous counterrotators.
We conclude that 
gaseous counterrotators are likely to have had a major merger 
(mass ratio close to or less than 2 to 1),
while the early-type galaxies with gas disks in the comparison sample seem not to have
experienced such mergers at the same rate as the counter-rotating objects, unless co-rotating major mergers produce less reduction in spin parameter.
We did not do a careful analysis 
to search for a surviving core of the merging galaxy, or 
to put limits on counter-rotating stars.
The MaNGA resolution may not be sufficient for such a study,
and further follow-up with higher resolution data would be useful.

The spin parameter of the lenticular counterrotators 
is also of interest.
As seen in the upper-right panel of Fig.~\ref{fig:spin_hist},
their spin parameters for the inner region 
are generally 
larger than those of the elliptical counterrotators, in agreement with what might be expected from a morphological difference,
but smaller than those of the lenticular comparison sample.
Their spin parameters for the outer region on the bottom right panel, 
on the other hand, appear to show two populations, one following 
that of the elliptical counterrotators, and the other that of the lenticular comparison sample. 
This double peak distribution of the spin parameter then may indicate 
the lenticular counterrotators have two distinctive formation origins:
one is the major merger, similar to the elliptical counterrotators,
and other minor mergers or gas accretion. The latter
would result in a relatively small change in the stellar kinematics.

\subsection{Star Formation in Gas Driven onto the Center} \label{sec:discussion_centralSF}

Gas acquisition in the opposite direction to the
spin of the host galaxy would drive the gas into the
center due to the collision of gas and resulting angular momentum loss \citep{Corsini2014,Chen2016,Jin2016,Bao2022}.
One possible mechanism to consume this gas at the center 
is star formation.
This process can be
efficiently produced if pre-existing gas
in the host galaxy is abundant.
\citet{Chen2016} pointed out this process from blue gaseous counterrotators showing a high central star formation rate. 
As we described in section \ref{sec:SFR}, 
star forming regions, if present, are centrally concentrated in the counterrotators,
while we could not see this concentration in the comparison sample.
This difference exists only for galaxies with sufficient star formation, i.e. those
whose star formation rate (SFR) is larger than 10\% of the star forming main sequence (SFMS).
The central concentration of the star forming regions 
is more significant in the lenticular counterrotators than in elliptical counterrotators.
A representative case of central star forming regions is 
a stellar counterrotators where some of its inner stars were
formed in the counter-rotating gas. We have some of these cases, as discussed in section \ref{sec:SCR}.
This inner young stellar component is
kinetically decoupled from the main stellar component, 
counter-rotating to the main stellar component and co-rotating with gas.

\citet{Chen2016} also claim that the central star formation 
can contribute to the central growth of the galaxy.
If this high star formation at the center induces
sufficient central growth, we would find a higher concentration of 
the stellar component compared to that of the corotators 
(the comparison sample).
But we did not find noticeable central growth in
the stellar counterrotators whose central star formation was strong enough 
to form a kinematically decoupled stellar component.
This might imply that the gas driven into the center
creates only a short episode of star formation with 
its contribution to the central growth of a galaxy limited.

The transportation of gas-inducing star formation at the center 
suggests outside-in quenching in the counterrotators.
However, we could not find evidence for the outside-in quenching signature in the analysis of the quenching ages, estimated from analysis of the stellar spectra, between the inner and outer regions
(e.g. that the outside-in quenching would be more common in the counterrotators).
This could arise from two possibilities. 
1) it is uncommon that a counterrotator has
gas driven into the center, and 
the central star formation happened in the past. 
For example, the gas acquisition onto a gas-(relatively)-free early-type 
would be unable to induce gas collision and angular momentum loss 
for the outside-in quenching.
2) If the quenching is made in a sufficiently short time scale,
we would not find a regional difference in the ages of the stellar populations.

\subsection{Central LIER Emission Feature} \label{sec:discussion_cLIER}

Based on the emission line ratios from the center, 
the fraction of cLIER (central low ionization emission line region) 
is 34\% for the gaseous counterrotators, higher than the 28\% for the comparison sample.
In addition, the fraction of cSF (central star formation) in the BPT diagram 
is 29\% in the gaseous counterrotators, lower than the 34\% for the comparison sample.
Meanwhile, the fractions of AGN and N/A (no classification due to weak emission lines) are similar in both samples.
The ionizing source of LIER galaxies could be weak AGN, shocks by outflow or 
formation process related to the gas acquisition, i.e. merger, 
or post-AGB stars, which would be formed at the center in the past.
An AGN is one of the potential ionizing sources
for gaseous counterrotators
because AGN feeding can be a process 
for the consumption of the gas driven into the center in the gaseous counterrotators.
Also, shocks resulting from the merger process are another probable ionizing source of the LINER/cLIER emissions in the gaseous counterrotators.
In particular, as seen in Section \ref{sec:spin} and \ref{sec:discussion_spin_parameter}, the elliptical counterrotators have spin parameters lower than the lenticular counterrotators.
And the fraction of LINER/cLIER is higher for the elliptical counterrotators
than for the lenticular counter-rotaors (section \ref{sec:BPT}).
Considering that a low spin parameter is induced by a major merger rather than a minor merger,
and a major merger can produce shocks ionizing the gas in the formation,
we favor a shock signature for the ionized gas.
However, the ionizing source could not be identified in detail in our analyses
due to the spectral resolution and IFU beam smearing effect 
(spatial resolution).
A further detailed analysis, with better resolution data,
would be useful to identify the ionizing source.

 \subsection{The Formation Origin of Gaseous Counterrotators} \label{sec:discussion_mergers}
 
Any formation scenario likely involves an outside origin for the counter-rotating gas component, 
given its angular momentum is drastically different from that of the stellar component.
An outside origin of gas would involve gas accretion from the circumgalactic medium (CGM) or merging with another galaxy.
The gas accretion can be both `episodic' or `continuous'
\citep{ThakarRyden1996,Thakar1997,ThakarRyden1998}.
Some recent studies reported that, from simulations and QSO absorption line observations, gas accretion from CGM to the ISM
is likely to produce co-rotation 
\citep{Ho2017, Ho2019}.
\citet{FraternaliBinney2006,Bizyaev2022} suggested that 
smooth accretion from the CGM may be responsible for the creation of`lags'
in the gas rotation in regular galaxies.
In this case, gas accretion can lower the rotation 
but not overwhelm the pre-existing rotation.
Since a large initial momentum
is required to produce the opposite rotation component
during the acquisition, gas accretion may be insufficient
to produce a counter-rotation, compared to a merging process, 
unless the galaxy onto which the gas is accreted was gas poor to begin with.
While the evidence for a low spin parameter discussed in Section \ref{sec:discussion_spin_parameter}
supports a merger origin for at least those systems with low spin parameter, 
we can not rule out a gas accretion origin for other systems,
in particular for gaseous counterrotators with high spin parameters.
The simulation of \citet{Brooks2009} suggested that cold gas accretion 
is more common in galaxy evolution compared to gas acquisition 
through galaxy merging interaction,
and \citet{Starkenburg2019} presented evidence for two low mass gaseous counterrotators 
formed by gas accretion.

For the merging scenario, the merger has to involve 
an infalling galaxy with abundant gas that
can sweep any pre-existing gas of the host galaxy. 
Thus, candidates for the infalling galaxy are likely to be 
a large galaxy with sufficient gas or a gas-rich dwarf galaxy.
The former would constitute a major merger,
and the latter a minor merger.
We prefer major mergers as the formation origin of the gaseous counterrotators for cases with low spin parameters in the outer region as well as the inner region. 
This is valid in most of the elliptical and about half of the lenticular gaseous counterrotators.
On the other hand, 
a major merger was often ruled out for the origin of the lenticular gaseous counterrotators
because such a merger in general is known to produce an elliptical galaxy, i.e. \citet{HernquistBernes1991} and \citet{Corsini2014}.
\citet{Puerari2001} reproduced major mergers resulting in massive counter-rotating galaxies with a disk component, such as NGC 4550 (which has two roughly equal stellar counter-rotating disks), though only a narrow range of the physical conditions were successful.
\citet{Zeng2021} showed that a major merger with a spiral-in falling can make 
a resulting galaxy with a disk component, and 
\citet{MartelRichard2020} shows a major merger can produce 
counter-rotating stars as well as gas with the massive star formation.
Further simulation studies would be useful to verify that major mergers can account for the number of gaseous counterrotators we find in early-type galaxies with low spin parameter.
We note that we are not arguing against minor mergers as the origin for some gaseous counterrotators.
As demonstrated in \citet{Smercina2023},
a minor merger is still a valid origin to produce a gaseous counterrotator
if the infalling galaxy is gas-rich.
In particular, the lenticular counterrotators with a relatively high spin parameter
might have a minor merger origin.

\section{Summary} \label{sec:summary}

\begin{itemize}

\item{We identify 303 gaseous counterrotators from
the whole sample of MaNGA galaxies, a fraction of 3\%.
The majority (about 80\%) of the gaseous counterrotators are early-types (defined here as elliptical or S0 galaxies), implying about 5.58\% of early-type galaxies in MaNGA show counter-rotation, 
and about 12.6\% if we consider only early-types with an overall rotating gas component.
They are mostly on the red sequence and the green valley regime in g-r color.
They are most frequent
in the intermediate stellar mass range around $10^{10}$ $M_{\odot}$.}

\item{To disentangle the morphological effects
from the effects of the counter-rotating gas, 
we compare the physical properties of the gaseous counterrotators and a comparison sample  within each morphological type. It turns out no significant difference is found for
the effective and 90\% light radii, their ratio, S\'ersic index, B/T, dust content and the detection rate of HI.
This proves that the morphological effects alone, not the gaseous counter-rotation, seems to be sufficient to explain
some characteristics of the gaseous counterrotators, namely  high central concentration of
stars within a galaxy (small radii and B/T ratio), low dust content, and low HI detection rate.}

\item{The gaseous counterrotators have three characteristics distinctive to the corotator comparison sample, namely galaxy rotation (low spin parameter for the stellar component), high concentration of star forming regions at the center if present, and
high LINER/cLIER galaxy fraction. 
The low spin parameter, implying weak rotational velocity compared to the velocity dispersion,
is confirmed for the region outside the effective radius as well as the inner region within the effective radius.
This might favor a major merger origin for the counter-rotating gas.
The high concentration of star forming regions at the center 
can be understood as the result of angular momentum loss in the gas component 
driving gas into the center and providing the fuel for star formation.
Gaseous counterrotators show a LINER/cLIER-type spectral signature in the center more frequently than comparison galaxies. 
This can be explained by AGN or shock ionization, possibly driven by galaxy interaction in the past. 
Given the limited spatial resolution of MaNGA, however,
we could not constrain a specific ionizing source.}

\item{We identify 33 potential merger candidates,
a minor fraction of the gaseous counterrotators, but this is with limited available data to constrain merger signatures. 
Their noticeable characteristics are a higher green valley galaxy fraction in the NUV-r color, 
and higher LINER/cLIER fractions,
while other physical properties are similar to those of the other gaseous counterrotators.}

\item{We discussed 25 cases of stellar as well as gaseous counterrotators, where the inner stellar component is counter-rotating to the major stellar component and co-rotating with the gas component. A large fraction of them have substantial star formation and/or very low spin parameters. They are considered as representative cases which had dramatic changes induced by the gas acquired in the opposite direction to the host galaxy spin.
This gas likely fueled massive star formation at the center
such that a new stellar population is rotating with the gas 
and counter-rotating to the main stellar component.}
 
\item{ We discussed three spirals whose spiral arms are strongly structured 
and show a `grand design' as in typical spiral galaxies.
Two of them have smooth light profiles 
in the center and spiral arms in the outskirts,
which might be in the transition phase from a spiral to a lenticular. 
The other spiral is in an ongoing interaction with its companion, 
a gas rich, 5 times massive galaxy in terms of stellar mass.}

\end{itemize}

\begin{acknowledgments}
We thank the NMSU Department of Astronomy for supporting the publication of this paper. 
DB is partly supported by RSCF grant 22-12-00080.

SDSS-IV is managed by the Astrophysical Research Consortium for the Participating Institutions of the SDSS Collaboration including the Brazilian Participation Group, the Carnegie Institution for Science, Carnegie Mellon University, the Chilean Participation Group, the French Participation Group, Harvard-Smithsonian Center for Astrophysics, Instituto de Astrof\'isica de Canarias, The Johns Hopkins University, Kavli Institute for the Physics and Mathematics of the Universe (IPMU) / University of Tokyo, Lawrence Berkeley National Laboratory, Leibniz Institut f\"ur Astrophysik Potsdam (AIP), Max-Planck-Institut f\"ur Astronomie (MPIA Heidelberg), Max-Planck-Institut f\"ur Astrophysik (MPA Garching), Max-Planck-Institut f\"ur Extraterrestrische Physik (MPE), National Astronomical Observatories of China, New Mexico State University, New York University, University of Notre Dame, Observat\'ario Nacional / MCTI, The Ohio State University, Pennsylvania State University, Shanghai Astronomical Observatory, United Kingdom Participation Group, Universidad Nacional Aut\'onoma de M\'exico, University of Arizona, University of Colorado Boulder, University of Oxford, University of Portsmouth, University of Utah, University of Virginia, University of Washington, University of Wisconsin, Vanderbilt University, and Yale University.
\end{acknowledgments}

\vspace{5mm}
\facilities{SDSS, MaNGA}

\software{IRAF, pPXF \citep{pPXF}
          }

\clearpage

\bibliographystyle{apj}

\end{document}